\documentclass[preprintnumbers]{revtex4}
\usepackage{amssymb}
\usepackage{amsmath}
\usepackage{graphicx}
\usepackage{dcolumn}
\usepackage{bm}
\usepackage{longtable}
\usepackage[latin1]{inputenc}
\usepackage{verbatim}

\newcommand{\ti}{\'{\i}}

\newcommand{\emp}{\begin{equation}}
\newcommand{\fin}{\end{equation}}
\newcommand{\empn}{\begin{equation*}}
\newcommand{\finn}{\end{equation*}}

\newcommand{\bea}{\begin{eqnarray}}
\newcommand{\eea}{\end{eqnarray}}
\newcommand{\eger}{\begin{gather}}
\newcommand{\fger}{\end{gather}}
\newcommand{\egn}{\begin{gather*}}
\newcommand{\fgn}{\end{gather*}}
\newcommand{\bit}{\begin{itemize}}
\newcommand{\eit}{\end{itemize}}

\newcommand{\lrp}[1]{\left(#1\right)}

\newcommand{\al}{\ensuremath{{\alpha}}}

\begin{document}

\title{Different kind of textures of Yukawa coupling matrices in the two
Higgs doublet model type III}
\author{A. E. C\'arcamo}
\email{aecarcamoh@unal.edu.co,antonio.carcamo@sns.it}
\affiliation{Departamento de F{\ti}sica, Universidad Nacional de Colombia,
Bogot\'a, Colombia\\
Scuola Normale Superiore Di Pisa, Pisa, Italy}
\author{R. Mart{\ti}nez}
\email{remartinezm@unal.edu.co}
\author{J.-Alexis Rodr{\ti}guez}
\email{jarodriguezl@unal.edu.co}
\affiliation{Departamento de F{\ti}sica, Universidad Nacional de Colombia\\
Bogot\'a, Colombia}
\date{}

\begin{abstract}
%\section*{\label{sec:level1}Abstract}\vspace{5cm}
The quark mass matrices ansatze proposed by Fritzsch, Du-Xing and
Fukuyama-Nishiura in the framework of the general two Higgs doublet model
are studied. The corresponding Yukawa matrices in the flavor basis in the
different cases considered are discussed. The corresponding
Cabbibo-Kobayashi-Maskawa matrix elements are computed in all cases and
compared with their experimental values. The complex phases of the ansatze
are taken into account and the CP violating phase $\delta$ is computed.
Finally, in order to observe the influence of the different kind of textures
of Yukawa coupling matrices considered, some phenomenology of two body
decays of the top quark, the lightest Higgs boson and the charged Higgs
boson is discussed.
\end{abstract}

\maketitle

%\preprint{APS/123-QED}
%\vspace{8cm}
%\title{Ansatz de masas de fermiones.}% Force line breaks with \\

%Bogot\'a, Colombia}%

%\vspace{0.45cm}

\section{Introduction}

Despite of all its success, the standard model (SM) of electroweak
interactions based on the $SU(3)_C\otimes SU(2)_L\otimes U(1)_Y$ gauge
symmetry has many unexplained features. Most of them are linked to the
fermionic sector, such as the origin of the fermion masses and the mixing
angles \cite{Fritzsch, Fx, Lavoura,Zhou}. These reasons lead to consider
that the SM is not the most fundamental theory of the basic interactions. It
should be considered as an effective theory that remains valid up to some
energy scale of the order of TeV and eventually it will be replaced by a
more fundamental theory. In the framework of the SM, the values of the
Yukawa couplings are parameterized in a phenomenological way. The quark
masses and mixing matrix are described by 10 free parameters: six quark
masses, three flavor mixing angles and one CP violating phase. It is, the
form of the Yukawa couplings is not well understood and neither their origin
nor the underlying principles, which it is known as the flavor problem.
Attempts to compute these 10 phenomenological parameters have been done
within the framework of the extensions of the SM including grand unification
theories, supersymmetric theories and superstring theories \cite%
{Giudice,Fukuyama, ibanez}. There are two basic approaches to study the
patterns of Yukawa couplings: one way is adopting the unification hypothesis
of matter multiplets and the second one is to assume a specific form of the
Yukawa couplings called texture \cite{ramond}. The two approaches are close
related because the mass matrices gotten using textures could be
incorporated into the grand unified theories \cite{Giudice,
Fukuyama-Nishiura}. The use of quark mass matrix textures is motivated by
the observed large hierarchies of the quark masses and the
Cabbibo-Kobayashi-Maskawa (CKM) matrix elements \cite{Xing,Cheng,Gupta,Branco}%
. Phenomenological quark mass matrices have been discussed from different
points of view. For example, quark mass matrix ansatze have been used in an
analogous way in the leptonic sector, trying to get the lepton mixing matrix
in order to explain the neutrino anomalies \cite%
{Fukuyama-Nishiura,Xing,Haba,Koide}. Also, new texture ansatze have been
proposed motivated by more precise results on the elements of the quark
mixing matrix \cite{Nishiura,Fritzsch,Cheng,Branco}. Quark and lepton mass
matrices have also been discussed in the context of the $SO\left(10\right)$
Grand Unification theories \cite{Fukuyama-Nishiura}. The understanding of
the discrete flavor symmetries hidden in such textures may be useful in the
knowledge of the underlying dynamics responsible for quark mass generation
and CP violation.

One possible simple extension of the SM is by adding a new Higgs doublet and
it is called the two Higgs Doublet Model (2HDM). This extension has the
following direct consequences: it increases the scalar spectrum and it gives
a more generic pattern of the Flavor Changing Neutral Currents (FCNC). FCNC
at tree level can be consider a problem that was solved in the earlier
versions of the two Higgs Doublet Model (2HDM type I and II) by imposing a
discrete symmetry that restricts each fermion to be coupled at most to one
Higgs doublet \cite{glashow}. But if the discrete symmetry is not imposed
then FCNC at tree level remains, it is the so-called two Higgs Doublet Model
type III. The 2HDM type III and its phenomenology have been extensively
studied in the literature \cite{Martinez}. In the 2HDM-III, for each type
quark, up or down type, there are two Yukawa couplings. One of the Yukawa
couplings is in charge of generating the quark masses and the other one
produces the flavor changing couplings at tree level. But in any case both
Higgs doublets have the same quantum numbers then if one ansatz is assumed
for the Yukawa coupling which generates the quark masses, is valid to
associate the same structure for the other Yukawa coupling which is
generating the flavor changing couplings, it is the main hypothesis of this
work.

In this manuscript, different kind of textures of Yukawa coupling matrices
are discussed in the framework of the 2HDM type III. In section \ref%
{sec:level1}, the principal features of the 2HDM type III are reviewed, then
in section \ref{sec:level2} the different kind of texture ansatze are
introduced. First of all, the Fritzsch ansatz and the Du-Xing ansatz are
reviewed. Then, the Fritzsch and Du-Xing ansatze are combined, searching a
better agreement with the experimental results. A brief review of the
Fukuyama-Nishiura mass matrix ansatz is presented and finally a new ansatz
is proposed. The new ansatz presents the feature that the top quark does not
have any mix and therefore it does not have any flavor changing neutral
process at tree level. The corresponding Yukawa coupling matrices in the
mass basis in the framework of the 2HDM type III are gotten for all the
ansatze discussed, it means the intensity of couplings that lead to FCNC
processes are obtained. Also, the numerical values of the corresponding CKM
matrix elements are obtained. In section \ref{sec:level3}, some
phenomenological aspects of the two body decays of the top quark, the
lightest Higgs boson and the charged Higgs boson are discussed. Finally in
section \ref{sec:level4} our conclusions are presented.

\section{\label{sec:level1} Yukawa interaction Lagrangian for the two Higgs
doublet model}

In the most general form of the two Higgs doublet model (2HDM), the
Lagrangian for the Yukawa interaction in the quark sector is given by \cite%
{Gunion},

\begin{align}
\begin{split}
-L_{Y}=\overline{{\ q}} ^0_L\eta^{U,0}\tilde{\phi}_1U^0_R+\overline{{\ q}}
^0_L\eta^{D,0}\phi_1D^0_R+\overline{{\ q}} ^0_L\xi^{U,0}\tilde{\phi}_2U^0_R+%
\overline{{\ q}} ^0_L\xi^{D,0}\phi_2D^0_R+h.c \label{a1}
\end{split}%
\end{align}
where $\eta^{f,0}$ and $\xi^{f,0}$ are the Yukawa interaction matrices being $%
\tilde{\phi}=i\tau_2\phi^{*}$, $f^0=U^0,D^0$, $q^0_L=\left(U^0,D^0\right)_L$
the quark doublet states in the interaction basis, and $U^0=(u^0,c^0,t^0)$, $%
D^0=(d^0,s^0,b^0)$ the quark interaction eigenstates. The two Higgs boson fields, after spontaneous symmetry breaking, have the following form:
\begin{equation}
\phi_k=%
\begin{pmatrix}
\phi^+_k \\
\frac1{\sqrt{2}}\left(v_k+\phi^0_k+i{\lambda} ^0_k\right)%
\end{pmatrix}%
\hspace{1cm}\mathit{with}\hspace{1cm}k=1,2 \label{a3}
\end{equation}
where $v_1$ and $v_2$ are the vacuum expectation values of the two Higgs
fields $\phi_1$ and $\phi_2$, respectively. There are not any extra discrete
symmetry acting on the Higgs fields, it means they can mix because they have
the same quantum numbers. Furthermore, both Higgs doublets can give mass to
down- and up-type quarks simultaneously. However, the Higgs fields can be
rotated to a new base where
\begin{equation}
\begin{pmatrix}
\phi_1^{\prime} \\
\phi_2^{\prime}%
\end{pmatrix}
=%
\begin{pmatrix}
\cos \beta & \sin \beta \\
-\sin \beta & \cos \beta%
\end{pmatrix}
\begin{pmatrix}
\phi_1 \\
\phi_2%
\end{pmatrix}%
\end{equation}
in such case $\tan \beta =v_2/v_1$ and the vacuum expectation values are $<
\phi_1^{\prime}>=v=\sqrt{v_1^2+v_2^2}$ and $<\phi^{\prime}_2>=0$. In this
article, we are not considering spontaneous CP symmetry breaking in the Higgs
sector. In the new mentioned base, the Yukawa Lagrangian can be written in
exactly the same way of equation (\ref{a1}), but the matrices $\eta^{f,0}$,
$\xi^{f,0}$ will be rotated to a new base
\begin{equation}
\begin{pmatrix}
\eta^{^{\prime}f,0} \\
\xi^{^{\prime}f,0}%
\end{pmatrix}
=
\begin{pmatrix}
\cos \beta & -\sin \beta \\
\sin \beta & \cos \beta%
\end{pmatrix}
\begin{pmatrix}
\eta^{f,0} \\
\xi^{f,0}%
\end{pmatrix}
.
\end{equation}
In general the quantum system will be in the prime base where only one
vacuum expectation value is different from zero \cite{Martinez}. We are
going to use this base for the Yukawa sector but now on we are going to omit
the prime symbol. In this new base, it can be shown that in the scalar
sector there are five scalar bosons,
\begin{equation}
\phi^{\pm}=H^\pm , \,\, \, \lambda_2^0=A^0 , \,\,\, \phi_1^0, \,\,\,
\phi_2^0,
\end{equation}
where $\phi_1^0$, $\phi_2^0$ rotate and in the mass eigenstates they are
\begin{equation}
\begin{pmatrix}
h^0 \\
H^0%
\end{pmatrix}%
=
\begin{pmatrix}
\cos \alpha & \sin \alpha \\
-\sin \alpha & \cos \alpha%
\end{pmatrix}
\begin{pmatrix}
\phi_1^0 \\
\phi_2^0%
\end{pmatrix}%
.
\end{equation}
Where the parameter $\alpha$ can be taken in the range $-\frac{\pi}{2}\leqslant\alpha\leqslant\frac{\pi}{2}$ \cite%
{Gunion}. Thus, from the scalar potential arises up five Higgs boson mass
eigenstates, they are $H^0$, $h^0$, $H^\pm$ and $A^0$.

From the Lagrangian's terms of expression (\ref{a1}), we get that the mass
matrices for fermions are given by,
\begin{equation}
M^f=\frac{v}{\sqrt{2}}\eta^{f,0} \, . \label{a16}
\end{equation}
And the Hermitian mass matrix M is diagonalized by a rotation matrix,
according to
\begin{equation}
V^{\dag}MV=O^TP^{\dag}MPO=O^T\widetilde{{\ M}} O=%
\begin{pmatrix}
\pm m_1 & 0 & 0 \\
0 & \mp m_2 & 0 \\
0 & 0 & m_3%
\end{pmatrix}
\label{a25}
\end{equation}
where $\widetilde{{\ M}} =P^{\dag}MP$ is a real symmetric mass matrix, $O$
is an orthogonal matrix, $P=diag\left(1,e^{-i\psi},e^{-i\left(\psi+\theta%
\right)}\right)$ and $m_1$, $m_2$, $m_3$ correspond to the fermion masses.
The upper signs are associated to the Fritzsch and Du-Xing ansatze, while
the lower ones are used for the Fukuyama-Nishiura ansatz (See section III).

The fermion mass eigenstates are related to the interaction eigenstates by
bi-unitary transformations \cite{Martinez},
\begin{equation}
U^0_L=V^U_LU_L,\hspace{1.5cm}U^0_R=V^U_RU_R,\hspace{1.5cm}D^0_L=V^D_LD_L,%
\hspace{1.5cm}D^0_R=V^D_RD_R \label{a30}
\end{equation}
and therefore the CKM matrix will be defined as
\begin{equation}
K=\left(V^U_L\right)^{\dag}V^D_L=O^T_UP_{UD}O_D,\hspace{1.5cm}\mathit{where}%
\hspace{1.5cm}P_{UD}=P^{\dag}_UP_D=diag\left(1,e^{i{\sigma}
},e^{i\tau}\right)\label{ckm}
\end{equation}
being ${\sigma} =\psi_U-\psi_D$ and $\tau=\psi_U-\psi_D+\theta_U-\theta_D $.
$\psi_{U,D}$ and $\theta_{U,D}$ are the phases of the mass matrices for the
up- and down-type quarks, respectively. Further, when the transition to the
mass eigenstates is performed in the Yukawa Lagrangian (\ref{a1}), the
following relationships corresponding to the ``rotated'' coupling matrices
are gotten \cite{Martinez},
\begin{equation}
M_U^{diag}=\frac{v}{\sqrt{2}}V_L^{U \dag} \eta^{U,0} V^U_R \hspace{1cm}
M_D^{diag}=\frac{v}{\sqrt{2}}V_L^{D \dag}\eta^{D,0} V^D_R \hspace{1cm}
\xi^U=V^{U \dag}_L\xi^{U,0} V^U_R,\hspace{1cm}\xi^D=V^{D \dag}_L\xi^{D,0}
V^D_R . \label{a31}
\end{equation}
Then eliminating $\eta^{f,0}$ and $\xi^{f,0}$ using equation (\ref{a31}),
the Yukawa Lagrangian corresponding to the quark sector for the two Higgs
Doublet Model type III takes the following form \cite{Martinez,Gunion},
\begin{eqnarray}
-L_Y &=& \overline{{\ U}} M^{diag} _UU+\overline{{\ D}} M^{diag}
_DD+\frac1{v}\overline{{\ U}} M^{diag} _UU\left(\cos\alpha H^0-\sin\alpha
h^0\right)+\frac1{\sqrt{2}}\overline{{\ U}} \left(\xi^UP_R+\left(\xi^U%
\right)^{\dag}P_L\right)U\left(\sin\alpha H^0+\cos\alpha h^0\right) \notag
\\
&+&\frac1{v}\overline{{\ D}} M^{diag} _DD\left(\cos\alpha H^0-\sin\alpha
h^0\right)+\frac1{\sqrt{2}}\overline{{\ D}} \left(\xi^DP_R+\left(\xi^D%
\right)^{\dag}P_L\right)D\left(\sin\alpha H^0+\cos\alpha h^0\right)-\frac{i}{%
v}\overline{{\ D}} M^{diag} _D{\gamma} _5DG^0_Z\hspace{0cm} \notag \\
&+&\frac{i}{v}\overline{{\ U}} M^{diag} _U{\gamma} _5UG^0_Z-\frac{i}{\sqrt{2}%
}\overline{{\ U}} \left(\xi^UP_R-\left(\xi^U\right)^{\dag}P_L\right)UA^0+%
\frac{i}{\sqrt{2}}\overline{{\ D}} \left(\xi^DP_R-\left(\xi^D\right)^{%
\dag}P_L\right)DA^0 \label{aly1} \\
&+&\frac{\sqrt{2}}{v}\overline{{\ U}} \left(KM^{diag} _DP_R-M^{diag}
_UKP_L\right)DG^+_W-\frac{\sqrt{2}}{v}\overline{{\ D}} \left(K^{%
\dag}M^{diag} _UP_R-M^{diag} _DK^{\dag}P_L\right)UG^-_W \notag \\
&+&\overline{{\ U}} \left(K\xi^DP_R-\left(\xi^U\right)^{\dag}KP_L\right)DH^+-%
\overline{{\ D}} \left(K^{\dag}\xi^UP_R-\left(\xi^D\right)^{\dag}K^{\dag}P_L%
\right)UH^- . \notag
\end{eqnarray}
Notice that if $\xi^U$ and $\xi^D$ vanish from this Yukawa Lagrangian, the
2HDM type II is reproduced. The new terms are leading to flavor changing
processes at tree level.

\section{\label{sec:level2} Different kind of ansatze}

In this section different mass matrices ansatze are presented. The Yukawa
coupling matrices in the mass basis are obtained, assuming that the Yukawa
coupling matrices have the same structure that the mass matrices. In the
framework of the 2HDM type III the flavor changing couplings at tree level
will depend on the ansatz parameters. The Fritzsch ansatz is the first one
reviewed because is the simplest one and it was the motivation for the
widely known and used Cheng-Sher ansatz in the 2HDM-III phenomenology. Then
the Du-Xing ansatz is discussed and it is followed by combinations of the
Fritzsch and the Du-Xing ansatze looking for the best fit of the CKM
elements. Furthermore, the Fukuyama-Nishiura ansatz is explored and a new
type of ansatz which is controlling the FCNC processes in the top quark
sector is proposed. It is worth to mention that phases are explicitly
involved since the mass matrices are hermitic, and then they can be written
always as $M =P \tilde M P^\dagger$, where $\tilde M$ is a real and
symmetric matrix and $P$ is a diagonal matrix which contains the phases \cite%
{Branco}.

\subsection{Fritzsch Ansatz (FA)}

For the three family case, a mass matrix ansatz proposed by Fritzsch has
been widely discussed\cite{Xing,Cheng,Fritzsch}, it is
\begin{equation}
M=%
\begin{pmatrix}
0 & D & 0 \\
D^* & 0 & B \\
0 & B^* & A%
\end{pmatrix}%
=%
\begin{pmatrix}
1 & 0 & 0 \\
0 & e^{-i\psi} & 0 \\
0 & 0 & e^{-i\left(\psi+\theta\right)}%
\end{pmatrix}%
\begin{pmatrix}
0 & |D| & 0 \\
|D| & 0 & |B| \\
0 & |B| & A%
\end{pmatrix}%
\begin{pmatrix}
1 & 0 & 0 \\
0 & e^{i\psi} & 0 \\
0 & 0 & e^{i\left(\psi+\theta\right)}%
\end{pmatrix}%
=P\widetilde{{\ M}} P^{\dag} . \label{a30}
\end{equation}
And motivated by the Fritzsch texture mass matrix, Cheng and Sher have
proposed an ansatz for the Yukawa flavor changing coupling matrices in the
interaction basis according to \cite{Xing,Cheng}:

\begin{equation}
\xi^{f,0}=\frac{\sqrt{2}}{v}%
\begin{pmatrix}
0 & dD & 0 \\
d D^* & 0 & b B \\
0 & b B^* & a A%
\end{pmatrix}%
=\frac{\sqrt{2}}{v}%
\begin{pmatrix}
1 & 0 & 0 \\
0 & e^{-i\psi} & 0 \\
0 & 0 & e^{-i\left(\psi+\theta\right)}%
\end{pmatrix}%
\begin{pmatrix}
0 & d|D| & 0 \\
d|D| & 0 & b|B| \\
0 & b|B| & aA%
\end{pmatrix}%
\begin{pmatrix}
1 & 0 & 0 \\
0 & e^{i\psi} & 0 \\
0 & 0 & e^{i\left(\psi+\theta\right)}%
\end{pmatrix}%
=P \tilde \xi^{f,0}P^{\dag} \label{a040}
\end{equation}
where $f=U,D$, the coefficients $a,b,d$ are of the order of one, and the
parameters $A$, $|B|$ and $|D|$ are given by,
\begin{equation}
A=m_3-m_2+m_1,\hspace{1.5cm}|D|=\left(\frac{m_1m_2m_3}{m_3-m_2+m_1}%
\right)^{1/2}, \notag
\end{equation}
\begin{equation}
|B|=\left(m_1m_2+m_3m_2-m_1m_3-\frac{m_1m_2m_3}{m_3+m_1-m_2}\right)^{1/2}.
\label{a35}
\end{equation}
But taking into account the mass hierarchy $m_{1}<<m_{2}<<m_{3}$, the
expressions (\ref{a35}) reduce to

\begin{equation}
A\approx m_3,\hspace{1.5cm}|B|\approx\sqrt{m_2m_3},\hspace{1.5cm}|D|\approx%
\sqrt{m_1m_2} . \label{a39}
\end{equation}
The rotation matrix which diagonalizes $\widetilde{{\ M}} $ is given by
\begin{equation}
O\simeq%
\begin{pmatrix}
1 & -\sqrt{\frac{m_1}{m_2}} & 0 \\
\sqrt{\frac{m_1}{m_2}} & 1 & \sqrt{\frac{m_2}{m_3}} \\
-\sqrt{\frac{m_1}{m_3}} & -\sqrt{\frac{m_2}{m_3}} & 1%
\end{pmatrix}
\label{rf}
\end{equation}
Then using the mass matrix (\ref{a30}) for up-type and down-type quarks and
equations (\ref{ckm}) and (\ref{rf}), the following expressions of the CKM
matrix elements are gotten

\begin{equation}
K_{ud}\simeq 1+e^{i{\sigma }}\sqrt{\frac{m_{u}m_{d}}{m_{c}m_{s}}}+e^{i\tau }%
\sqrt{\frac{m_{u}m_{d}}{m_{t}m_{b}}},\hspace{1.5cm}K_{us}\simeq e^{i{\sigma }%
}\sqrt{\frac{m_{u}}{m_{c}}}-\sqrt{\frac{m_{d}}{m_{s}}}+e^{i\tau }\sqrt{\frac{%
m_{u}m_{s}}{m_{t}m_{b}}}, \notag
\end{equation}%
\begin{equation}
K_{ub}\simeq \sqrt{\frac{m_{u}}{m_{c}}}\left( e^{i{\sigma }}\sqrt{\frac{m_{s}%
}{m_{b}}}-e^{i\tau }\sqrt{\frac{m_{c}}{m_{t}}}\right) ,\hspace{1cm}%
K_{cd}\simeq e^{i{\sigma }}\sqrt{\frac{m_{d}}{m_{s}}}-\sqrt{\frac{m_{u}}{%
m_{c}}}+e^{i\tau }\sqrt{\frac{m_{c}m_{d}}{m_{t}m_{b}}},
\end{equation}%
\begin{equation}
K_{cs}\simeq e^{i{\sigma }}+\sqrt{\frac{m_{u}m_{d}}{m_{c}m_{s}}}+e^{i\tau }%
\sqrt{\frac{m_{c}m_{s}}{m_{t}m_{b}}},\hspace{1.5cm}K_{cb}\simeq e^{i{\sigma }%
}\sqrt{\frac{m_{s}}{m_{b}}}-e^{i\tau }\sqrt{\frac{m_{c}}{m_{t}}}, \notag
\label{v2f}
\end{equation}%
\begin{equation}
K_{td}\simeq \sqrt{\frac{m_{d}}{m_{s}}}\left( e^{i{\sigma }}\sqrt{\frac{m_{c}%
}{m_{t}}}-e^{i\tau }\sqrt{\frac{m_{s}}{m_{b}}}\right) ,\hspace{1cm}%
K_{ts}\simeq e^{i{\sigma }}\sqrt{\frac{m_{c}}{m_{t}}}-e^{i\tau }\sqrt{\frac{%
m_{s}}{m_{b}}},\hspace{1cm}K_{tb}\simeq e^{i\tau }+e^{i{\sigma }}\sqrt{\frac{%
m_{c}m_{s}}{m_{t}m_{b}}}\,. \notag \label{v3f}
\end{equation}%
In order to get the best agreement of the $\bigl|K_{us}\bigr|$ and $\bigl|%
K_{cd}\bigr|$ elements with their experimental values, we use ${\sigma }%
=81.57^{\circ }$ and $\tau =-11.78^{\circ }$. Hence $\psi _{U}=\psi
_{D}+81.57^{\circ}$ and $\theta _{U}=\theta_{D}-93.35^{\circ}$. The numerical
values shown in table \ref{tab:table2} for the CKM matrix elements are obtained using the mass values listed in appendix A. From these results, a considerable disagreement of the $\bigl|K_{ub}\bigr|$, $\bigl|K_{cb}\bigr|$, $\bigl|K_{td}\bigr|$ and $\bigl|K_{ts}\bigr|$ elements with their experimental values is noticed. On the other hand, the CP violating phase $\delta $ for the Fritzsch ansatz
is obtained, $\delta =86.61^{\circ }$ by using the expression \cite{Branco}:
\begin{equation}
\sin \delta =\frac{\lrp{1-\bigl|K_{ub}\bigr|^{2}}J}{\bigl|K_{ud}K_{us}K_{ub}K_{cb}K_{tb}\bigr|} \, ,
\end{equation}
where $J$ is the Jarlskog invariant given by $J=\Im m\lrp{K_{us}K_{cb}K_{ub}^{\ast}K_{cs}^{\ast}}$ \cite{Branco,Jarlskog}. This CP phase $\delta$ presents a
discrepancy of about $30.54\%$ with respect to the experimental value $%
\delta =60.16^{\circ }\pm 14^{\circ }$.\\

On other hand, the Yukawa coupling matrices in the mass basis are given by
\cite{Cheng}
\begin{equation}
\widetilde{{\ \xi }}^{U,D}=O_{U,D}^{T}\widetilde{{\ \xi }}^{U,D,0}O_{U,D}\,,
\label{a65}
\end{equation}%
where $\widetilde{{\ \xi }}^{U,D,0}=P^{\dag }\xi ^{U,D,0}P$. Therefore,
taking into account the mass hierarchy and using the expression $\xi
^{U,D}=PO_{U,D}^{T}\tilde{\xi}^{U,D,0}O_{U,D}P^{\dag }$, the Yukawa coupling
matrices in the mass basis are given by\\

\begin{equation}
\xi^U\simeq\frac{\sqrt{2}}{v}%
\begin{pmatrix}
\left(2d^{\left(U\right)} -2b^{\left(U\right)} +a^{\left(U\right)} \right)m_u
& \left(a^{\left(U\right)} +d^{\left(U\right)} -2b^{\left(U\right)} \right)%
\sqrt{m_um_c}e^{i\psi_U} & \left(b^{\left(U\right)} -a^{\left(U\right)}
\right)\sqrt{m_um_t}^{i\left(\psi_U+\theta_U\right)} \\
\left(a^{\left(U\right)} +d^{\left(U\right)} -2b^{\left(U\right)} \right)%
\sqrt{m_um_c}e^{-i\psi_U} & \left(a^{\left(U\right)} -2b^{\left(U\right)}
\right)m_c & \left(b^{\left(U\right)} -a^{\left(U\right)} \right)\sqrt{m_cm_t%
}e^{i\theta_U} \\
\left(b^{\left(U\right)} -a^{\left(U\right)} \right)\sqrt{m_um_t}%
e^{-i\left(\psi_U+\theta_U\right)} & \left(b^{\left(U\right)}
-a^{\left(U\right)} \right)\sqrt{m_cm_t}e^{-i\theta_U} & a^{\left(U\right)}
m_t%
\end{pmatrix}
\, , \label{a71}
\end{equation}

\begin{equation}
\xi^D\simeq\frac{\sqrt{2}}{v}%
\begin{pmatrix}
\left(2d^{\left(D\right)} -2b^{\left(D\right)} +a^{\left(D\right)} \right)m_d
& \left(a^{\left(D\right)} +d^{\left(D\right)} -2b^{\left(D\right)} \right)%
\sqrt{m_dm_s}e^{i\psi_D} & \left(b^{\left(D\right)} -a^{\left(D\right)}
\right)\sqrt{m_dm_b}^{i\left(\psi_D+\theta_D\right)} \\
\left(a^{\left(D\right)} +d^{\left(D\right)} -2b^{\left(D\right)} \right)%
\sqrt{m_dm_s}e^{-i\psi_D} & \left(a^{\left(D\right)} -2b^{\left(D\right)}
\right)m_s & \left(b^{\left(D\right)} -a^{\left(D\right)} \right)\sqrt{m_sm_b%
}e^{i\theta_D} \\
\left(b^{\left(D\right)} -a^{\left(D\right)} \right)\sqrt{m_dm_b}%
e^{-i\left(\psi_D+\theta_D\right)} & \left(b^{\left(D\right)}
-a^{\left(D\right)} \right)\sqrt{m_sm_b}e^{-i\theta_D} & a^{\left(D\right)}
m_b%
\end{pmatrix}
\, , \notag
\end{equation}
equation (\ref{a71}) reproduces the expression obtained by Cheng and Sher
\cite{Cheng} when the complex phases vanish. Further, we can see that the
hierarchy structure of the Yukawa coupling matrices in the mass basis are
dominated by the fermion masses and the $\left(\xi^{U,D}\right)_{ij}$
elements are proportional to $\sqrt{m_im_j}$, which is the usual Cheng-Sher
ansatz \cite{Cheng}.

\begin{table}[h]
\begin{tabular}{|c|c|c|c|c|c|c|c|}
\hline
& FA & DXA & DXFA & FDXA & $\tilde{X}$A & FNA & NTA \\ \hline
$\bigl|K_{ud}\bigr|$ & 1.00189 & 1.00189 & 1.00189 & 1.00189 & 1.00189 &
0.97467 & 1.00051 \\ \hline
$\bigl|K_{us}\bigr|$ & 0.22720 & 0.22720 & 0.22720 & 0.22720 & 0.22720 &
0.22361 & 0.22720 \\ \hline
$\bigl|K_{ub}\bigr|$ & 0.00797 & 0.00090 & 0.00734 & 0.00359 & 0.00181 &
0.00308 & 0.00190 \\ \hline
$\bigl|K_{cd}\bigr|$ & 0.22710 & 0.22721 & 0.22710 & 0.22710 & 0.22718 &
0.22341 & 0.22720 \\ \hline
$\bigl|K_{cs}\bigr|$ & 1.00144 & 1.00194 & 1.00146 & 1.00144 & 1.00178 &
0.97381 & 1.00051 \\ \hline
$\bigl|K_{cb}\bigr|$ & 0.15512 & 0.01746 & 0.14277 & 0.06986 & 0.03562 &
0.04221 & 0.03197 \\ \hline
$\bigl|K_{td}\bigr|$ & 0.03559 & 0.00401 & 0.03275 & 0.01603 & 0.00817 &
0.01002 & 0.00735 \\ \hline
$\bigl|K_{ts}\bigr|$ & 0.15512 & 0.01746 & 0.14277 & 0.06986 & 0.03562 &
0.04112 & 0.03203 \\ \hline
$\bigl|K_{tb}\bigr|$ & 0.99955 & 1.00004 & 0.99956 & 0.99955 & 0.99988 &
0.99910 & 1 \\ \hline
$\delta $ & $86.61^{\circ }$ & $86.66^{\circ }$ & $86.62^{\circ }$ & $%
86.66^{\circ }$ & $82.18^{\circ }$ & $-87.55^{\circ }$ & $80.68^{\circ }$ \\
\hline\hline
\end{tabular}%
\caption{The CKM matrix elements and CP violating phase $\protect\delta $
for the different ansatze considered.}
\label{tab:table2}
\end{table}

\subsection{Du-Xing Ansatz (DXA)}

In order to accommodate all the current data of quark masses and mixing
angles in the framework of texture-zeros, the following ansatz for the mass
matrix was suggested \cite{Xing,Cheng,Fritzsch},

\begin{equation}
M=%
\begin{pmatrix}
0 & D & 0 \\
D^* & C & B \\
0 & B^* & A%
\end{pmatrix}%
=P\widetilde{{\ M}} P^{\dag}\hspace{1.0cm}\mathit{with} \hspace{1.0cm} \bigl|%
A\bigr|\gg\bigl|B\bigr|, \bigl|C\bigr|\gg\bigl|D\bigr|
\label{dxb}
\end{equation}
where $A\simeq m_3$, $|B|\simeq|C|\simeq m_2$, and $|D|\simeq\sqrt{m_1m_2}$.
The rotation matrix for this case is \cite{Cheng}:

\begin{equation}
O\simeq%
\begin{pmatrix}
1 & \sqrt{\frac{m_1}{m_2}} & 0 \\
-\sqrt{\frac{m_1}{m_2}} & 1 & \frac{m_2}{m_3} \\
\frac{\sqrt{m_1m_2}}{m_3} & -\frac{m_2}{m_3} & 1%
\end{pmatrix}
\, . \label{b016}
\end{equation}

And the CKM matrix elements obtained from the mass matrix ansatz given in (\ref{dxb}) are \cite{Xing}

\begin{equation}
K_{ud}\simeq 1+e^{i{\sigma} }\sqrt{\frac{m_um_d}{m_cm_s}}+e^{i\tau}\frac{%
\sqrt{m_um_cm_dm_s}}{m_tm_b},\hspace{1.5cm}K_{us}\simeq\sqrt{\frac{m_d}{m_s}}%
-e^{i{\sigma} }\sqrt{\frac{m_u}{m_c}}-e^{i\tau}\frac{m_s\sqrt{m_um_c}}{m_tm_b%
} \, , \notag
\end{equation}
\begin{equation}
K_{ub}\simeq\sqrt{\frac{m_u}{m_c}}\left(\frac{m_c}{m_t}e^{i\tau}-\frac{m_s}{%
m_b}e^{i{\sigma} }\right),\hspace{1.5cm}K_{cd}\simeq\sqrt{\frac{m_u}{m_c}}%
-e^{i{\sigma} }\sqrt{\frac{m_d}{m_s}}-e^{i\tau}\frac{m_c\sqrt{m_dm_s}}{m_tm_b%
} \, , \label{vz1}
\end{equation}
\begin{equation}
K_{cs}\simeq e^{i{\sigma} }+\sqrt{\frac{m_um_d}{m_cm_s}}+\frac{m_cm_s}{m_tm_b%
}e^{i\tau},\hspace{1.5cm}K_{cb}\simeq\frac{m_s}{m_b}e^{i{\sigma} }-\frac{m_c%
}{m_t}e^{i\tau} \, , \notag \label{vz2}
\end{equation}
\begin{equation}
K_{td}\simeq\sqrt{\frac{m_d}{m_s}}\left(\frac{m_s}{m_b}e^{i\tau}-\frac{m_c}{%
m_t}e^{i{\sigma} }\right),\hspace{1.5cm}K_{ts}\simeq\frac{m_c}{m_t}e^{i{%
\sigma} }-\frac{m_s}{m_b}e^{i\tau},\hspace{1.5cm}K_{tb}\simeq e^{i\tau}+%
\frac{m_cm_s}{m_tm_b}e^{i{\sigma} } \, . \notag \label{vz3}
\end{equation}

The best agreement of the $\bigl|K_{us}\bigr|$ and $\bigl|K_{cd}\bigr|$
elements with their experimental values is gotten when ${\sigma }%
=81.09^{\circ }$ and $\tau =29.97^{\circ }.$ Then $\psi _{U}=\psi
_{D}+81.09^{\circ }$ and $\theta _{U}=\theta _{D}-51.\,\allowbreak 12^{\circ
}$. The numerical values of the CKM matrix elements are shown in table \ref{tab:table2}. There are a considerable disagreement between the magnitudes of the $\bigl|K_{ub}\bigr|$, $\bigl|K_{cb}\bigr|$, $\bigl|K_{td}\bigr|$ and $\bigl|K_{ts}\bigr|$ elements and their experimental values. On the other hand, the obtained CP violating phase for the Du-Xing ansatz is $\delta =86.66^{\circ }$. This result has an inconsistency with the experimental value \cite{PDG} of about $30.58\%$.\newline

\begin{table}[tbp]
\begin{tabular}{|c|c|c|c|c|c|c|c|}
\hline
& FA & DXA & DXFA & FDXA & $\tilde{X}$A & FNA & NTA \\ \hline
$\bigl|K_{ud}\bigr|$ & 2.80 & 2.80 & 2.80 & 2.80 & 2.80 & 0.09 & 2.67 \\
\hline
$\bigl|K_{us}\bigr|$ & 0.00 & 0.00 & 0.00 & 0.00 & 0.00 & 1.60 & 0.00 \\
\hline
$\bigl|K_{ub}\bigr|$ & 50.32 & 341.21 & 46.03 & 10.30 & 118.31 & 28.72 &
108.64 \\ \hline
$\bigl|K_{cd}\bigr|$ & 0.00 & 0.05 & 0.00 & 0.00 & 0.03 & 1.65 & 0.04 \\
\hline
$\bigl|K_{cs}\bigr|$ & 2.84 & 2.89 & 2.84 & 2.84 & 2.88 & 0.09 & 2.75 \\
\hline
$\bigl|K_{cb}\bigr|$ & 72.79 & 141.69 & 70.44 & 38.58 & 18.50 & 0.00 & 32.01
\\ \hline
$\bigl|K_{td}\bigr|$ & 77.13 & 103.16 & 75.15 & 49.21 & 0.39 & 18.80 & 10.77
\\ \hline
$\bigl|K_{ts}\bigr|$ & 73.18 & 138.25 & 70.86 & 40.44 & 16.81 & 1.20 & 29.91
\\ \hline
$\bigl|K_{tb}\bigr|$ & 0.04 & 0.09 & 0.05 & 0.04 & 0.08 & 0.00 & 0.09 \\
\hline
$\delta $ & 30.54 & 30.58 & 30.55 & 30.58 & 26.79 & 168.72 & 25.43 \\
\hline\hline
\end{tabular}%
\caption{Error percentages of the CKM matrix elements and CP violating phase
$\protect\delta $ for the different ansatze considered. The experimental
values used are those reported by \protect\cite{PDG}}
\label{tab:table3}
\end{table}
The table \ref{tab:table3} shows the error percentages of the CKM matrix
elements and CP violating phase $\delta $ for the different ansatze. The
deviations of the CKM matrix elements obtained from the Du-Xing ansatz are
higher than the corresponding to the Fritzsch ansatz. Then, the Fritzsch ansatz
leads to a better prediction of the CKM matrix elements than the Du-Xing ansatz. About the CP violating phase, the Fritzsch ansatz leads to approximately the same prediction of this phase than the obtained from the Du-Xing ansatz.\\

On other hand, the Yukawa coupling matrices in the mass basis are
\begin{equation}
\xi^U\simeq\frac{\sqrt{2}}{v}%
\begin{pmatrix}
\left(c^{\left(U\right)} -2d^{\left(U\right)} \right)m_u &
\left(d^{\left(U\right)} -c^{\left(U\right)} \right)\sqrt{m_um_c}e^{i\psi_U}
& \left(a^{\left(U\right)} -b^{\left(U\right)} \right)\sqrt{m_um_c}%
e^{i\left(\psi_U+\theta_U\right)} \\
\left(d^{\left(U\right)} -c^{\left(U\right)} \right)\sqrt{m_um_c}e^{-i\psi_U}
& c^{\left(U\right)} m_c & \left(b^{\left(U\right)} -a^{\left(U\right)}
\right)m_ce^{i\theta_U} \\
\left(a^{\left(U\right)} -b^{\left(U\right)} \right)\sqrt{m_um_c}%
e^{-i\left(\psi_U+\theta_U\right)} & \left(b^{\left(U\right)}
-a^{\left(U\right)} \right)m_ce^{-i\theta_U} & a^{\left(U\right)} m_t%
\end{pmatrix}
\label{a710}
\end{equation}

\begin{equation}
\xi^D\simeq\frac{\sqrt{2}}{v}%
\begin{pmatrix}
\left(c^{\left(D\right)} -2d^{\left(D\right)} \right)m_d &
\left(d^{\left(D\right)} -c^{\left(D\right)} \right)\sqrt{m_dm_s}e^{i\psi_D}
& \left(a^{\left(D\right)} -b^{\left(D\right)} \right)\sqrt{m_dm_s}%
e^{i\left(\psi_D+\theta_D\right)} \\
\left(d^{\left(D\right)} -c^{\left(D\right)} \right)\sqrt{m_dm_s}e^{-i\psi_D}
& c^{\left(D\right)} m_s & \left(b^{\left(D\right)} -a^{\left(D\right)}
\right)m_se^{i\theta_D} \\
\left(a^{\left(D\right)} -b^{\left(D\right)} \right)\sqrt{m_dm_s}%
e^{-i\left(\psi_D+\theta_D\right)} & \left(b^{\left(D\right)}
-a^{\left(D\right)} \right)m_se^{-i\theta_U} & a^{\left(D\right)} m_b%
\end{pmatrix}
\label{a711}
\end{equation}
using the Du-Xing ansatz (\ref{dxb}) according to $\xi^{U,D}=P\widetilde{{\
\xi}} ^{U,D}P^{\dag}$. Then, the Du-Xing ansatz lead to the same structure
of the Yukawa couplings for the first and the second generation fermions.
And the coupling to the third generation fermions in the expressions
(\ref{a710}) and (\ref{a711}) are weaker than the predicted by the Fritzsch ansatz.\newline

\subsection{Combination of the Fritzsch and the Du-Xing ansatz}

\subsubsection{\textbf{The Du-Xing ansatz for the up sector and the Fritzsch
ansatz for the down sector (DFXA)}}

By using the Fritzsch and the Du-Xing ansatze for the down- and up-type
quarks respectively, the following expressions corresponding to the CKM
matrix elements are obtained,

\begin{equation}
K_{ud}\simeq 1-e^{i{\sigma }}\sqrt{\frac{m_{d}m_{u}}{m_{s}m_{c}}}-e^{i\tau }%
\frac{\sqrt{m_{u}m_{c}m_{d}}}{m_{t}\sqrt{m_{b}}},\hspace{1.5cm}K_{us}\simeq -%
\sqrt{\frac{m_{d}}{m_{s}}}-e^{i{\sigma }}\sqrt{\frac{m_{u}}{m_{c}}}-e^{i\tau
}\frac{\sqrt{m_{u}m_{c}m_{s}}}{m_{t}\sqrt{m_{b}}}\,, \notag \label{g5}
\end{equation}%
\begin{equation}
K_{ub}\simeq e^{i\tau }\frac{\sqrt{m_{u}m_{c}}}{m_{t}}-e^{i{\sigma }}\sqrt{%
\frac{m_{u}m_{s}}{m_{c}m_{b}}},\hspace{1.5cm}K_{cd}\simeq e^{i{\sigma }}%
\sqrt{\frac{m_{d}}{m_{s}}}+\sqrt{\frac{m_{u}}{m_{c}}}+e^{i\tau }\frac{m_{c}}{%
m_{t}}\sqrt{\frac{m_{d}}{m_{b}}}\,, \label{g6}
\end{equation}%
\begin{equation}
K_{cs}\simeq e^{i{\sigma }}+e^{i\tau }\frac{m_{c}}{m_{t}}\sqrt{\frac{m_{s}}{%
m_{b}}}-\sqrt{\frac{m_{d}m_{u}}{m_{s}m_{c}}},\hspace{1.5cm}K_{cb}\simeq e^{i{%
\sigma }}\sqrt{\frac{m_{s}}{m_{b}}}-e^{i\tau }\frac{m_{c}}{m_{t}}\,, \notag
\label{g7}
\end{equation}%
\begin{equation}
K_{td}\simeq e^{i{\sigma }}\frac{m_{c}}{m_{t}}\sqrt{\frac{m_{d}}{m_{s}}}%
-e^{i\tau }\sqrt{\frac{m_{d}}{m_{b}}},\hspace{1.5cm}K_{ts}\simeq e^{i{\sigma
}}\frac{m_{c}}{m_{t}}-e^{i\tau }\sqrt{\frac{m_{s}}{m_{b}}},\hspace{1.5cm}%
K_{tb}\simeq e^{i\tau }+e^{i{\sigma }}\frac{m_{c}}{m_{t}}\sqrt{\frac{m_{s}}{%
m_{b}}}\,. \notag \label{g8}
\end{equation}

With the aim to get a better agreement of the $\bigl|K_{ud}\bigr|$
and $\bigl|K_{cd}\bigr|$ elements with their experimental values we obtain ${%
\sigma }=81.09^{\circ }$ and $\tau =29.97^{\circ }$ , and then $\psi
_{U}=\psi _{D}+90^{\circ }$ and $\theta _{U}=\theta _{D}-51.12^{\circ}$. The
numerical values of the CKM matrix elements for this case are shown in Table \ref{tab:table2}. In this case, there is good agreement of the CKM matrix elements with
their experimental values, with the exception of the $\bigl|K_{ub}\bigr|$, $\bigl|K_{cb}\bigr|$, $\bigl|K_{td}\bigr|$
and $\bigl|K_{ts}\bigr|$ elements, which present discrepancies of about $%
46.03\%$, $70.44\%$, $75.15\%$ and $70.86\%$, respect to their corresponding experimental values. On the other hand, we obtain that the CP violating phase is $\delta =86.62^{\circ }$. This result presents a discrepancy of about $30.55\%$.\\

About the Yukawa coupling matrices in the mass basis for this case, they are
\begin{equation}
\xi ^{U}\simeq \frac{\sqrt{2}}{v}%
\begin{pmatrix}
\left( c^{\left( U\right) }-2d^{\left( U\right) }\right) m_{u} & \left(
d^{\left( U\right) }-c^{\left( U\right) }\right) \sqrt{m_{u}m_{c}}e^{i\psi
_{U}} & \left( a^{\left( U\right) }-b^{\left( U\right) }\right) \sqrt{%
m_{u}m_{c}}e^{i\left( \psi _{U}+\theta _{U}\right) } \\
\left( d^{\left( U\right) }-c^{\left( U\right) }\right) \sqrt{m_{u}m_{c}}%
e^{-i\psi _{U}} & c^{\left( U\right) }m_{c} & \left( b^{\left( U\right)
}-a^{\left( U\right) }\right) m_{c}e^{i\theta _{U}} \\
\left( a^{\left( U\right) }-b^{\left( U\right) }\right) \sqrt{m_{u}m_{c}}%
e^{-i\left( \psi _{U}+\theta _{U}\right) } & \left( b^{\left( U\right)
}-a^{\left( U\right) }\right) m_{c}e^{-i\theta _{U}} & a^{\left( U\right)
}m_{t}%
\end{pmatrix}%
\,, \label{zf8}
\end{equation}

\begin{equation}
\xi^D\simeq\frac{\sqrt{2}}{v}%
\begin{pmatrix}
\left(2d^{\left(D\right)} -2b^{\left(D\right)} +a^{\left(D\right)} \right)m_d
& \left(a^{\left(D\right)} +d^{\left(D\right)} -2b^{\left(D\right)} \right)%
\sqrt{m_dm_s}e^{i\psi_D} & \left(b^{\left(D\right)} -a^{\left(D\right)}
\right)\sqrt{m_dm_b}^{i\left(\psi_D+\theta_D\right)} \\
\left(a^{\left(D\right)} +d^{\left(D\right)} -2b^{\left(D\right)} \right)%
\sqrt{m_dm_s}e^{-i\psi_D} & \left(a^{\left(D\right)} -2b^{\left(D\right)}
\right)m_s & \left(b^{\left(D\right)} -a^{\left(D\right)} \right)\sqrt{m_sm_b%
}e^{i\theta_D} \\
\left(b^{\left(D\right)} -a^{\left(D\right)} \right)\sqrt{m_dm_b}%
e^{-i\left(\psi_D+\theta_D\right)} & \left(b^{\left(D\right)}
-a^{\left(D\right)} \right)\sqrt{m_sm_b}e^{-i\theta_D} & a^{\left(D\right)}
m_b%
\end{pmatrix}
\, . \notag \label{zf9}
\end{equation}

\subsubsection{\textbf{The Fritzsch ansatz for the up quark sector and the
Du-Xing ansatz for the down quark sector (FDXA)}}

In this section we use the Fritzsch and the Du-Xing ansatze for the up-type
and down-type quarks respectively. It is contrary to the DXFA case discussed before. The expressions for the CKM matrix elements are:

\begin{equation}
K_{ud}\simeq 1-e^{i{\sigma} }\sqrt{\frac{m_dm_u}{m_sm_c}}-e^{i\tau}\frac{%
\sqrt{m_um_dm_s}}{m_b\sqrt{m_t}},\hspace{1.5cm}K_{us}\simeq\sqrt{\frac{m_d}{%
m_s}}+e^{i{\sigma} }\sqrt{\frac{m_u}{m_c}}+e^{i\tau}\frac{m_s}{m_b}\sqrt{%
\frac{m_u}{m_t}} \, , \notag \label{g06}
\end{equation}
\begin{equation}
K_{ub}\simeq e^{i{\sigma} }\frac{m_s}{m_b}\sqrt{\frac{m_u}{m_c}}-e^{i\tau}%
\sqrt{\frac{m_u}{m_t}},\hspace{1.5cm}K_{cd}\simeq-e^{i{\sigma} }\sqrt{\frac{%
m_d}{m_s}}-\sqrt{\frac{m_u}{m_c}}-e^{i\tau}\frac{\sqrt{m_cm_dm_s}}{m_b\sqrt{%
m_t}} \, , \label{g6}
\end{equation}
\begin{equation}
K_{cs}\simeq e^{i{\sigma} }+e^{i\tau}\frac{m_s}{m_b}\sqrt{\frac{m_c}{m_t}}-%
\sqrt{\frac{m_dm_u}{m_sm_c}},\hspace{1.5cm}K_{cb}\simeq e^{i{\sigma} }\frac{%
m_s}{m_b}-e^{i\tau}\sqrt{\frac{m_c}{m_t}} \, , \notag \label{h5}
\end{equation}
\begin{equation}
K_{td}\simeq e^{i\tau}\frac{\sqrt{m_dm_s}}{m_b}-e^{i{\sigma} }\sqrt{\frac{%
m_dm_c}{m_sm_t}},\hspace{1.5cm}K_{ts}\simeq e^{i{\sigma} }\sqrt{\frac{m_c}{%
m_t}}-e^{i\tau}\frac{m_s}{m_b},\hspace{1.5cm}K_{tb}\simeq e^{i\tau}+e^{i{%
\sigma} }\frac{m_s}{m_b}\sqrt{\frac{m_c}{m_t}} \, . \notag \label{h6}
\end{equation}

In order to get the best agreement of the $\bigl|K_{us}\bigr|$ and $\bigl|%
K_{cd}\bigr|$ elements with their experimental values, we get ${\sigma }%
=-98.93^{\circ }$ and $\tau =52.19^{\circ }$. And $\psi _{U}=\psi_{D}-98.93^{\circ }$ and $\theta _{U}=\theta_{D}+151.12^{\circ }$ are found. Then according to the previous results, the numerical values are presented in Table \ref{tab:table2}. In this case, there is good agreement of the CKM matrix elements with their experimental values, with the exception of the $\bigl|K_{cb}\bigr|$, $\bigl|K_{td}\bigr|$, $\bigl|K_{ts}\bigr|$ elements, which present discrepancies of about $38.58\%$, $49.21\%$, and $40.44\%$, respect to their corresponding experimental values. On the other hand, we obtain that the CP violating phase for this case is $\delta=86.66^{\circ}$. This result is deviated $30.58\%$ from the experimental value.\\

In the Table \ref{tab:table3}, the error percentages of the CKM matrix
elements are lower than the corresponding to the $\bigl|K_{ij}\bigr|$ elements
given by the DFXA case. This is an indication that when the prescription FDXA
is used, we obtain a better agreement with the experimental values of the CKM
matrix elements than in the combinations considered before.\\

On the other hand, the following Yukawa coupling matrices in the mass basis
are obtained,
\begin{equation}
\xi^U\simeq\frac{\sqrt{2}}{v}%
\begin{pmatrix}
\left(2d^{\left(U\right)} -2b^{\left(U\right)} +a^{\left(U\right)} \right)m_u
& \left(a^{\left(U\right)} +d^{\left(U\right)} -2b^{\left(U\right)} \right)%
\sqrt{m_um_c}e^{i\psi_U} & \left(b^{\left(U\right)} -a^{\left(U\right)}
\right)\sqrt{m_um_t}^{i\left(\psi_U+\theta_U\right)} \\
\left(a^{\left(U\right)} +d^{\left(U\right)} -2b^{\left(U\right)} \right)%
\sqrt{m_um_c}e^{-i\psi_U} & \left(a^{\left(U\right)} -2b^{\left(U\right)}
\right)m_c & \left(b^{\left(U\right)} -a^{\left(U\right)} \right)\sqrt{m_cm_t%
}e^{i\theta_U} \\
\left(b^{\left(U\right)} -a^{\left(U\right)} \right)\sqrt{m_um_t}%
e^{-i\left(\psi_U+\theta_U\right)} & \left(b^{\left(U\right)}
-a^{\left(U\right)} \right)\sqrt{m_cm_t}e^{-i\theta_U} & a^{\left(U\right)}
m_t
\end{pmatrix}
\label{fz1}
\end{equation}

\begin{equation}
\xi^D\simeq\frac{\sqrt{2}}{v}%
\begin{pmatrix}
\left(c^{\left(D\right)} -2d^{\left(D\right)} \right)m_d &
\left(d^{\left(D\right)} -c^{\left(D\right)} \right)\sqrt{m_dm_s}e^{i\psi_D}
& \left(a^{\left(D\right)} -b^{\left(D\right)} \right)\sqrt{m_dm_s}%
e^{i\left(\psi_D+\theta_D\right)} \\
\left(d^{\left(D\right)} -c^{\left(D\right)} \right)\sqrt{m_dm_s}e^{-i\psi_D}
& c^{\left(D\right)} m_s & \left(b^{\left(D\right)} -a^{\left(D\right)}
\right)m_se^{i\theta_D} \\
\left(a^{\left(D\right)} -b^{\left(D\right)} \right)\sqrt{m_dm_s}%
e^{-i\left(\psi_D+\theta_D\right)} & \left(b^{\left(D\right)}
-a^{\left(D\right)} \right)m_se^{-i\theta_U} & a^{\left(D\right)} m_b%
\end{pmatrix}
\label{fz2}
\end{equation}

\subsection{Combination of different assignments in the Du-Xing ansatz($%
(\tilde X A)$}

It is known that taking $M^{diag} =diag\left(-m_1, m_2, m_3\right)$,
the following relations between the components of the $\widetilde{{\ M}} $
matrix given in (\ref{dxb}) are satisfied \cite{Xing},
\begin{equation}
C+A=-m_1+m_2+m_3,\hspace{1.2cm}CA-|B|^2-|D|^2=-m_1m_2+m_2m_3+m_3m_1,\hspace{%
1.2cm}A|D|^2=m_1m_2m_3 . \label{cn4}
\end{equation}
The assumptions $A\gg |B|$ and $|C|\gg |D|$ define the assignment type A and
the assumption $C=m_2$ define the assignment type B. For the assignment type
B, the parameters are
\begin{equation}
A=m_3-m_1,\hspace{1.5cm}|D|=\sqrt{\frac{m_1m_2m_3}{m_3-m_1}},\hspace{1.5cm}
|B|=\sqrt{\frac{m_1m_3\left(m_3-m_2-m_1\right)}{m_3-m_1}} , \label{cn5}
\end{equation}
and the rotation matrix $O$ is \cite{Fukuyama-Nishiura}:
\begin{equation}
O \simeq
\begin{pmatrix}
1 & \sqrt{\frac{m_1}{m_2}} & \sqrt{\frac{m_2m^2_1}{m^3_3}} \\
-\sqrt{\frac{m_1}{m_2}} & 1 & \sqrt{\frac{m_1}{m_3}} \\
\sqrt{\frac{m^2_1}{m_2m_3}} & -\sqrt{\frac{m_1}{m_3}} & 1%
\end{pmatrix}%
\hspace{1.5cm}\mathit{for}\hspace{1.5cm}m_3\gg m_2\gg m_1 .\hspace{0.6cm}
\label{cn6}
\end{equation}

Then, the following expressions for the CKM matrix elements are obtained,

\begin{equation}
K_{ud}\simeq 1+e^{i{\sigma} }\sqrt{\frac{m_um_d}{m_cm_s}}+e^{i\tau}\frac{m_d%
}{m_t}\sqrt{\frac{m_um_c}{m_sm_b}},\hspace{1.5cm}K_{us}\simeq\sqrt{\frac{m_d%
}{m_s}}-e^{i{\sigma} }\sqrt{\frac{m_u}{m_c}}-e^{i\tau}\frac{\sqrt{m_um_cm_d}%
}{m_t\sqrt{m_b}} \, , \notag
\end{equation}
\begin{equation}
K_{ub}\simeq\sqrt{\frac{m_sm^2_d}{m^3_b}}-e^{i{\sigma} }\sqrt{\frac{m_um_d}{%
m_cm_b}}+e^{i\tau}\frac{\sqrt{m_um_c}}{m_t},\hspace{1.5cm}K_{cd}\simeq\sqrt{%
\frac{m_u}{m_c}}-\frac{m_c}{m_t}\sqrt{\frac{m^2_d}{m_sm_b}}e^{i\tau}-e^{i{%
\sigma} }\sqrt{\frac{m_d}{m_s}} \, , \label{cn8}
\end{equation}
\begin{equation}
K_{cs}\simeq e^{i{\sigma} }+\sqrt{\frac{m_um_d}{m_cm_s}}+e^{i\tau}\frac{m_c}{%
m_t}\sqrt{\frac{m_d}{m_b}},\hspace{1.5cm}K_{cb}\simeq e^{i{\sigma} }\sqrt{%
\frac{m_d}{m_b}}-\frac{m_c}{m_t}e^{i\tau}+\sqrt{\frac{m_um_sm^2_d}{m_cm^3_b}}
\, , \notag \label{cn9}
\end{equation}
\begin{equation}
K_{td}\simeq e^{i\tau}\sqrt{\frac{m^2_d}{m_sm_b}}-e^{i{\sigma} }\frac{m_c}{%
m_t}\sqrt{\frac{m_d}{m_s}},\hspace{1.5cm}K_{ts}\simeq e^{i{\sigma} }\frac{m_c%
}{m_t}-e^{i\tau}\sqrt{\frac{m_d}{m_b}},\hspace{1.5cm}K_{tb}\simeq e^{i\tau}+%
\frac{m_c}{m_t}\sqrt{\frac{m_d}{m_b}}e^{i{\sigma} } \, . \notag
\label{cn10}
\end{equation}

In this case, ${\sigma }=81.09^{\circ }$, $\tau =-98.91^{\circ }$ are
obtained to get a good experimental agreement of the $\bigl|K_{us}\bigr|$
and $\bigl|K_{cd}\bigr|$ elements with their experimental values and then $%
\psi _{U}=\psi _{D}+81.09^{\circ }$, $\theta _{U}=\theta _{D}-180^{\circ }$.
The obtained numerical values for the CKM matrix elements are in Table \ref{tab:table2}. The CP violating phase is $\delta =82.18^{\circ }$ presenting
a discrepancy of $26.79\%$. From Table \ref{tab:table3}, notice that the
error percentages of the magnitudes of the CKM matrix elements are lower
than $10.00\%$ with the exception of the corresponding to the $\bigl|K_{ub}%
\bigr|$, $\bigl|K_{cb}\bigr|$ and $\bigl|K_{ts}\bigr|$ elements which are equal to $118.31\%$, $18.50\%$ and $16.81\%$, respectively. \newline

The corresponding $\xi^U$ and $\xi^D$ Yukawa coupling matrices in the mass
basis for this case are given by,

\begin{equation}
\xi^U\simeq\frac{\sqrt{2}}{v}%
\begin{pmatrix}
\left(c^{\left(U\right)} -2d^{\left(U\right)} \right)m_u &
\left(d^{\left(U\right)} -c^{\left(U\right)} \right)\sqrt{m_um_c}e^{i\psi_U}
& a^{\left(U\right)} \frac{m_u}{m_c}\sqrt{m_cm_t}e^{i\left(\psi_U+\theta_U%
\right)} \\
\left(d^{\left(U\right)} -c^{\left(U\right)} \right)\sqrt{m_um_c}e^{-i\psi_U}
& c^{\left(U\right)} m_c & \left(b^{\left(U\right)} -a^{\left(U\right)}
\right)\sqrt{m_um_t}e^{i\theta_U} \\
a^{\left(U\right)} \frac{m_u}{m_c}\sqrt{m_cm_t}e^{-i\left(\psi_U+\theta_U%
\right)} & \left(b^{\left(U\right)} -a^{\left(U\right)} \right)\sqrt{m_um_t}%
e^{-i\theta_U} & a^{\left(U\right)} m_t%
\end{pmatrix}
\, , \label{a071}
\end{equation}

\begin{equation}
\xi^D\simeq\frac{\sqrt{2}}{v}%
\begin{pmatrix}
\left(c^{\left(D\right)} -2d^{\left(D\right)} \right)m_d &
\left(d^{\left(D\right)} -c^{\left(D\right)} \right)\sqrt{m_dm_s}e^{i\psi_D}
& a^{\left(D\right)} \frac{m_d}{m_s}\sqrt{m_sm_b}e^{i\left(\psi_D+\theta_D%
\right)} \\
\left(d^{\left(D\right)} -c^{\left(D\right)} \right)\sqrt{m_dm_s}e^{-i\psi_D}
& c^{\left(D\right)} m_s & \left(b^{\left(D\right)} -a^{\left(D\right)}
\right)\sqrt{m_dm_b}e^{i\theta_D} \\
a^{\left(U\right)} \frac{m_d}{m_s}\sqrt{m_sm_b}e^{-i\left(\psi_D+\theta_D%
\right)} & \left(b^{\left(D\right)} -a^{\left(D\right)} \right)\sqrt{m_dm_b}%
e^{-i\theta_D} & a^{\left(D\right)} m_b%
\end{pmatrix}
\, . \notag
\end{equation}

From the expressions (\ref{a071}), we observe that the Yukawa couplings to
the first and the second generation fermions have the same form that the
predicted by the Fritzsch and the Du-Xing ansatze. The couplings to the
third generation fermions are stronger than the predicted by the Du-Xing
ansatz and weaker than the obtained by the Cheng and Sher ansatz. This fact
has consequences in the phenomenology as we will show in section IV.\newline

\subsection{Fukuyama-Nishiura ansatz (FNA)}

Recently the following mass matrix ansatz has been proposed for quarks and
leptons \cite{Fukuyama-Nishiura}:
\begin{equation}
\widetilde{{\ M}} =%
\begin{pmatrix}
0 & A & A \\
A & B & C \\
A & C & B%
\end{pmatrix}
\label{ab1}
\end{equation}
This form has been originally used for leptons (neutrinos) in order to
reproduce a nearly bi-maximal lepton mixing. Moreover the mass matrix $%
\widetilde{{\ M}} $ is diagonalized by a rotation matrix according to:
\begin{equation}
O^T\widetilde{{\ M}} O=diag\left(-m_1,m_2,m_3\right) \label{ab2}
\end{equation}
where $m_1$, $m_2$ and $m_3$ correspond to the fermion masses with $m_1\ll
m_2\ll m_3$. By taking into account that $tr\left(\widetilde{{\ M}} \right)$
and $det\left(\widetilde{{\ M}} \right)$ are invariants and using the
eigenvalues equation for $\widetilde{{\ M}} $, we get:
\begin{equation}
A=\pm\sqrt{\frac{m_1m_2}{2}},\hspace{1.5cm}B=\frac1{2}\left(m_2+m_3-m_1%
\right),\hspace{1.5cm}C=-\frac1{2}\left(m_3-m_2+m_1\right). \label{ab8}
\end{equation}
The case where $B-C$ takes its maximum value corresponds to the assignment
type C. The assignment type D is obtained by exchanging $m_2$ and $m_3$ in
the already mentioned type C (\ref{ab8}). For the assignment type D,
\begin{equation}
A=\pm\sqrt{\frac{m_3m_1}{2}},\hspace{1.5cm}B=\frac1{2}\left(m_3+m_2-m_1%
\right),\hspace{1.5cm}C=\frac1{2}\left(m_3-m_2-m_1\right) . \label{ab083}
\end{equation}
The assignments type C and D are used for down-type and up-type quarks,
respectively. The rotation matrices which diagonalize $\widetilde{{\ M}} $
for the assignments type C and D are,
\begin{equation}
O=%
\begin{pmatrix}
c & s & 0 \\
-\frac{s}{\sqrt{2}} & \frac{c}{\sqrt{2}} & -\frac1{\sqrt{2}} \\
-\frac{s}{\sqrt{2}} & \frac{c}{\sqrt{2}} & \frac1{\sqrt{2}}%
\end{pmatrix}%
,\hspace{1.5cm}O^{\prime}=%
\begin{pmatrix}
c^{\prime} & 0 & s^{\prime} \\
-\frac{s^{\prime}}{\sqrt{2}} & -\frac1{\sqrt{2}} & \frac{c^{\prime}}{\sqrt{2}%
} \\
-\frac{s^{\prime}}{\sqrt{2}} & \frac1{\sqrt{2}} & \frac{c^{\prime}}{\sqrt{2}}%
\end{pmatrix}
\label{ab22}
\end{equation}
respectively, where
\begin{equation}
c=\cos\varphi =\sqrt{\frac{m_2}{m_2+m_1}},\hspace{1.5cm}s=\sin\varphi =\sqrt{%
\frac{m_1}{m_1+m_2}} \, , \notag \label{ab23}
\end{equation}
\begin{equation}
c^{\prime}=\cos\varphi ^{\prime}=\pm\sqrt{\frac{m_3}{m_1+m_3}},\hspace{1.5cm}%
s^{\prime}=\sin\varphi ^{\prime}=\pm\sqrt{\frac{m_1}{m_1+m_3}} \, . \notag
\label{ab24}
\end{equation}

In order to get a good agreement of $\frac{1}{9}\sum_{i=1}^{9}\bigl|K_{ij}%
\bigr|$ with their experimental values, we get ${\sigma }=-180^{\circ }$ and
${\tau }=4.84^{\circ }$. From this result, the numerical values
corresponding to the updated magnitudes of the CKM matrix elements for the
Fukuyama-Nishiura ansatz are obtained \cite{Fukuyama-Nishiura} (see Table %
\ref{tab:table2}).

From Table III, there are a very good agreement of the CKM matrix elements with
their experimental values since the corresponding error percentages are
lower than $2\%$. The exceptions are the $\bigl|K_{ub}\bigr|$ and $\bigl|K_{td}\bigr|$ elements, which present a discrepancies of about $28.72\%$ and $18.80\%$ respect to its corresponding experimental magnitudes. It is also important to point out that the Fukuyama-Nishiura ansatz leads to a much better prediction of the CKM matrix elements than the obtained from the Fritzsch and the Du-Xing ansatze. The $CP $ violating phase $\delta =-87.55^{\circ }$ in this case is inconsistent to its experimental value because it presents a discrepancy of about $168.72\%$. Furthermore $\theta _{U}=\theta _{D}+184.84$ and $\psi _{U}=\psi _{D}-4.84^{\circ }$ are obtained.

Finally, the Yukawa coupling matrices in the mass basis are given by
\begin{eqnarray}
\xi^U &\simeq& \frac{\sqrt{2}}{2 v}%
\begin{pmatrix}
\left(b^{\left(U\right)} +c^{\left(U\right)} -4a^{\left(U\right)} \right)m_u
& 0 & f^{(U)} \sqrt{m_um_t}e^{i\left(\psi_U+\theta_U\right)} \\
0 & \left(b^{\left(U\right)} -c^{\left(U\right)}
\right)m_t+\left(b^{\left(U\right)} +c^{\left(U\right)} \right)m_c & 0 \\
f^{(U)} \sqrt{m_um_t}e^{-i\left(\psi_U+\theta_U\right)} & 0 &
\left(b^{\left(U\right)} +c^{\left(U\right)} \right)m_t%
\end{pmatrix}
\, , \notag \\
\xi^D&\simeq&\frac{\sqrt{2}}{2 v}%
\begin{pmatrix}
2 \left(b^{\left(D\right)}-2a^{\left(D\right)}\right)m_d &
\left(b^{\left(D\right)}-c^{\left(D\right)}\right)\sqrt{\frac{m_d}{m_s}}%
m_be^{i\psi_D} & 0 \\
\left(b^{\left(D\right)}-c^{\left(D\right)}\right)\sqrt{\frac{m_d}{m_s}}%
m_be^{-i\psi_D} & \left(b^{\left(D\right)}-c^{\left(D\right)}\right)m_b+%
\left(b^{\left(D\right)}+c^{\left(D\right)}\right)m_s & 0 \\
0 & 0 & \left(b^{\left(D\right)}+c^{\left(D\right)}\right)m_b%
\end{pmatrix}
\, . \label{ab49}
\end{eqnarray}
where $f^{(U)} =\left(2a^{\left(U\right)} +b^{\left(U\right)}
-c^{\left(U\right)} \right)$. From the above expressions, the couplings
between the neutral Higgs field $h^0$ and the quarks pairs $u$-$c$, $c$-$t$,
$d$-$b$ and $s$-$b$ are vanishing, which it is meaning that there are no
flavor changing neutral currents involving terms like $h^0\overline{{\ u}} c$%
, $h^0\overline{{\ c}} t$, $h^0\overline{{\ d}} b$, $h^0\overline{{\ s}} b$.

\subsection{Non-mixing Top quark Ansatz (NTA)}

We propose the following mass matrix ansatz for up-type quarks
\begin{equation}
\widetilde{{\ M}} =%
\begin{pmatrix}
B & C & 0 \\
C & D & 0 \\
0 & 0 & A%
\end{pmatrix}
\label{e0}
\end{equation}
which is diagonalized by a rotation matrix $O$ given by
\begin{equation}
O=%
\begin{pmatrix}
c_\phi & -s_\phi & 0 \\
s_\phi & c_\phi & 0 \\
0 & 0 & 1%
\end{pmatrix}
\label{e3}
\end{equation}
according to $O^T \widetilde{{\ M}} O =diag(m_1,-m_2,m_3)$. In this case the
condition $\tan \phi=\sqrt{\frac{m_2}{m_3}}$ has been imposed and using the
expressions
\begin{equation}
det\widetilde{{\ M}} =A\left(BD-C^2\right),\hspace{1.5cm}tr\widetilde{{\ M}}
=B+D=-m_2+m_1 \, , \notag \label{e5}
\end{equation}
it is obtained that

\begin{equation}
A=m_3,\hspace{1.5cm}B=\frac{m_1m_3-m^2_2}{m_2+m_3},\hspace{1.5cm}C=\frac{%
m_1+m_2}{m_2+m_3}\sqrt{m_2m_3},\hspace{1.5cm}D=m_2\frac{m_1-m_3}{m_2+m_3} \,
, \label{e6}
\end{equation}
and therefore, the rotation matrix is

\begin{equation}
O=%
\begin{pmatrix}
\sqrt{\frac{m_3}{m_2+m_3}} & -\sqrt{\frac{m_2}{m_2+m_3}} & 0 \\
\sqrt{\frac{m_2}{m_2+m_3}} & \sqrt{\frac{m_3}{m_2+m_3}} & 0 \\
0 & 0 & 1%
\end{pmatrix}
\, . \label{e7}
\end{equation}
By using the mass matrix ansatz given by (\ref{e0}) for the up-type quarks
and the assignment type B of the Du-Xing ansatz for down-type quarks (\ref{dxb}), the following CKM matrix elements are obtained

\begin{equation}
K_{ud}\simeq c_U-s_U\sqrt{\frac{m_d}{m_s}}e^{i{\sigma} },\hspace{1.5cm}%
K_{us}\simeq c_U\sqrt{\frac{m_d}{m_s}}+s_Ue^{i{\sigma} },\hspace{1.5cm}%
K_{ub}\simeq c_U\sqrt{\frac{m^2_dm_s}{m^3_b}}+s_U\sqrt{\frac{m_d}{m_b}}e^{i{%
\sigma} } \label{e10}
\end{equation}
\begin{equation}
K_{cd}\simeq -s_U-c_U\sqrt{\frac{m_d}{m_s}}e^{i{\sigma} },\hspace{1.5cm}%
K_{cs}\simeq c_Ue^{i{\sigma} }-s_U\sqrt{\frac{m_d}{m_s}},\hspace{1.5cm}%
K_{cb}\simeq c_U\sqrt{\frac{m_d}{m_b}}e^{i{\sigma} }-s_U\sqrt{\frac{m^2_dm_s%
}{m^3_b}} \notag \label{e11}
\end{equation}
\begin{equation}
K_{td}\simeq\sqrt{\frac{m^2_d}{m_sm_b}}e^{i\tau},\hspace{1.5cm}K_{ts}\simeq
-e^{i\tau}\sqrt{\frac{m_d}{m_b}},\hspace{1.5cm}K_{tb}\simeq e^{i\tau} \, ,
\notag \label{e12}
\end{equation}
where:
\begin{equation}
c_U=\sqrt{\frac{m_t}{m_c+m_t}},\hspace{1.5cm}s_U=\sqrt{\frac{m_c}{m_c+m_t}}
\, . \notag \label{e13c}
\end{equation}
\\

In order to get the best agreement of the $\bigl|K_{us}\bigr|$ and $\bigl|%
K_{cd}\bigr|$ elements with their experimental values, we obtain ${\sigma
}=-99.24^{\circ }$. On the other hand, by adjusting the $\tau$ parameter in
order to get the best agreement of $\Im m\lrp{K_{ts}}$ with its experimental
value, $\tau=90.00^{\circ}$ is gotten. Then, we obtain the magnitudes of the
CKM matrix elements shown in Table \ref{tab:table2}. Moreover, for this case
we have $\psi _{U}=\psi _{D}-99.24^{\circ }$, $\theta _{U}=\theta
_{D}+189.24^{\circ}$. And for the $CP$ violating phase we obtain $\delta =80.68^{\circ }$. This result roughly agrees with the experimental value, presenting a discrepancy of about $25.43\%$.\\

The Table \ref{tab:table3} shows the error percentages of the magnitudes of
the Cabbibo-Kobayashi-Maskawa matrix elements and CP violating phase $\delta$
respect to their experimental values. In this Table we observe that the
error percentages of the magnitudes of the CKM matrix elements are lower
than $11.00\%$ with the exception of the corresponding to the $\bigl|K_{ub}%
\bigr|$, $\bigl|K_{cb}\bigr|$ and $\bigl|K_{ts}\bigr|$ elements which are equal to $108.64\%$, $32.01\%$ and $29.91\%$, respectively. For this reason, the
Cabbibo-Kobayashi-Maskawa matrix obtained by using the above prescription
exhibits a good agreement with the experimental results.\\

According to the mass matrix structure given by (\ref{e0}), we propose the
following ansatz for the Yukawa coupling matrix in the flavor basis:

\begin{equation}
\xi^{U,0}=\frac{\sqrt{2}}{v}%
\begin{pmatrix}
b^{\left(U\right)} B_U & c^{\left(U\right)} C_Ue^{i\psi_U} & 0 \\
c^{\left(U\right)} C_Ue^{-i\psi_U} & d^{\left(U\right)} D_U & 0 \\
0 & 0 & a^{\left(U\right)} A_U%
\end{pmatrix}%
\, . \label{e16}
\end{equation}

Hence, when the new ansatz and type B assignment of the Du-Xing ansatz are
used for the quarks $U$ and $D$ respectively, the Yukawa couplings matrices
in the mass basis are given by:

\begin{equation}
\xi^U=\frac{\sqrt{2}}{v}%
\begin{pmatrix}
b^{\left(U\right)} m_u & \left(c^{\left(U\right)} -d^{\left(U\right)} \right)%
\frac{m_c}{m_t}\sqrt{m_cm_t}e^{i\psi_U} & 0 \\
\left(c^{\left(U\right)} -d^{\left(U\right)} \right)\frac{m_c}{m_t}\sqrt{%
m_cm_t}e^{-i\psi_U} & -d^{\left(U\right)} m_c & 0 \\
0 & 0 & a^{\left(U\right)} m_t%
\end{pmatrix}
\, , \label{e21}
\end{equation}

\begin{equation}
\xi^D\simeq\frac{\sqrt{2}}{v}%
\begin{pmatrix}
\left(c^{\left(D\right)} -2d^{\left(D\right)} \right)m_d &
\left(d^{\left(D\right)} -c^{\left(D\right)} \right)\sqrt{m_dm_s}e^{i\psi_D}
& a^{\left(D\right)} \frac{m_d}{m_s}\sqrt{m_sm_b}e^{i\left(\psi_D+\theta_D%
\right)} \\
\left(d^{\left(D\right)} -c^{\left(D\right)} \right)\sqrt{m_dm_s}e^{-i\psi_D}
& c^{\left(D\right)} m_s & \left(b^{\left(D\right)} -a^{\left(D\right)}
\right)\sqrt{m_dm_b}e^{i\theta_D} \\
a^{\left(U\right)} \frac{m_d}{m_s}\sqrt{m_sm_b}e^{-i\left(\psi_D+\theta_D%
\right)} & \left(b^{\left(D\right)} -a^{\left(D\right)} \right)\sqrt{m_dm_b}%
e^{-i\theta_D} & a^{\left(D\right)} m_b%
\end{pmatrix}
\, . \notag \label{a072}
\end{equation}

From the previous expressions, the couplings between the neutral Higgs field
$h^0$ and the quarks pairs $u$-$t$, $c$-$t$ are vanishing, which implies
that there are no flavor changing neutral currents involving terms of the
form $h^0\overline{{\ u}} t$, $h^0\overline{{\ c}} t$.

\section{\label{sec:level3}Two body decays of the top quark, $h^0$ and $%
H^\pm $}

Consider the top quark decay to an up-type quark plus the lightest neutral
Higgs boson $h^0$. The interest on FCNC is expected to increase since this
issue will be examined at both the LHC and ILC, where they hope to reach
sensibilities of the order of $B(t \to qh^0)\gtrsim 10^{-5}$. In the
framework of the SM, the branching fractions are strongly suppressed $B(t
\to ch^0)\sim 10^{-15}$ and $B(t \to u h^0) \sim 10^{-17}$. In the framework
of the 2HDM type III an enhancement is expected due to the FCNC presence at
tree level. The decay width takes the form
\begin{equation}
\Gamma(t \to qh^0)=\frac{G_F m_t}{4 \sqrt{2} \pi} \left(1-\frac{m_h^2}{m_t^2}%
\right)^2 \vert \lambda_{tq}^U \vert^2 \cos^2\alpha
\end{equation}
where $\lambda_{tq}^U$ is the coupling $tqh^0$. Therefore, there are two
possible decays: $t \to uh^0$ and $t\to ch^0$. In figure 1, the branching
fraction $B(t \to c h^0)$ as a function of the lightest neutral Higgs boson
mass for the Fritzsch ansatz is plotted and orders of $10^{-3}$ are gotten.
In this Figure 1, a decrease of the branching ratio with the increase of the
Higgs mass is observed, but notice that the branching ratio raise up
increasing the magnitudes of the coefficients $a^{\left(U\right)}-b^{\left(U%
\right)}$ of the coupling $tch^0$. And the branching ratio decrease with the
reduction of the mixing angle $\alpha$ between the neutral Higgs fields $h^0$
and $H^0$. These coefficients and the mixing angle ${\alpha} $ were taken to
be equal to $0.75$, $1.00$ and $\frac{\pi}{4}$, $\frac{\pi}{15}$, $\frac{4\pi%
}{9}$, respectively. The branching fraction $B(t \to u h^0)$ is two orders
of magnitude lower than $B(t \to c h^0)$ and its behavior is quite similar.%
\newline

In order to consider the different ansatze, the following ratio is useful,
\begin{equation}
R^{\left(A/F\right)}=\frac{B\left(t\to qh^0\right)_A}{B\left(t\to
qh^0\right)_F}=\biggl \vert \frac{\lambda_{tq,A}^U}{\lambda_{tq,F}^U} \biggr
\vert^2
\end{equation}
it is respect to the Fritzsch ansatz (F), with $q=u,c$ and $%
A=DXA,FDXA,DXFA,\tilde XA, FNA, NTA$. It does not depend on $m_t$, $m_h$ or $%
\cos \alpha$, it depends on the explicit form of the masses in the different
kind of ansatze discussed. This ratio is typically of the order of $10^{-3}$
but in the case of the FNA and NTA such a kind of top quark decay width is
zero because the top quark is completely decoupled of the two lighter
generations.\\

\begin{figure}[tbp]
\begin{center}
% GNUPLOT: LaTeX picture with Postscript
\begingroup%
  \makeatletter%
  \newcommand{\GNUPLOTspecial}{%
    \@sanitize\catcode`\%=14\relax\special}%
  \setlength{\unitlength}{0.1bp}%
\begin{picture}(4680,3024)(0,0)%
\special{psfile=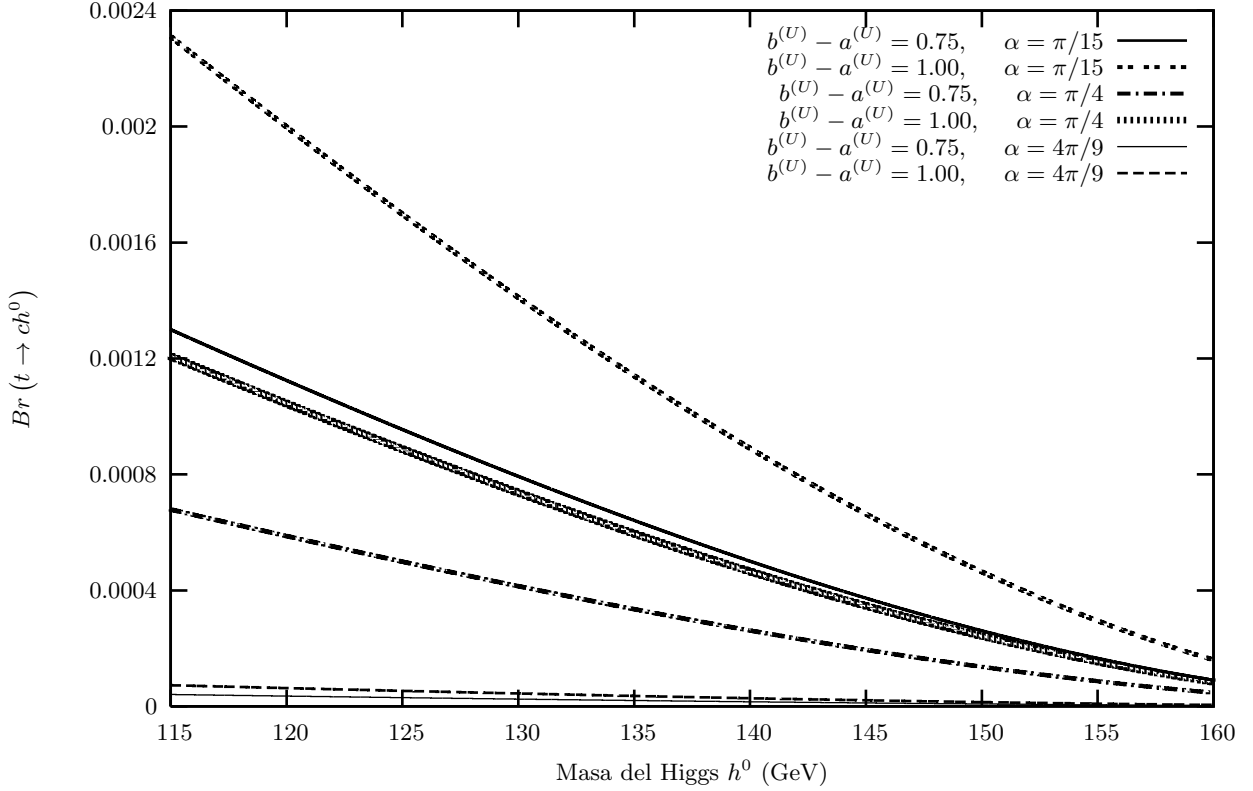 llx=0 lly=0 urx=468 ury=302 rwi=4680}
\put(4117,2311){\makebox(0,0)[r]{$b^{\lrp{U}}-a^{\lrp{U}}=1.00$,\hspace{0.5cm}$\al=4\pi/9$}}%
\put(4117,2411){\makebox(0,0)[r]{$b^{\lrp{U}}-a^{\lrp{U}}=0.75$,\hspace{0.5cm}$\al=4\pi/9$}}%
\put(4117,2511){\makebox(0,0)[r]{$b^{\lrp{U}}-a^{\lrp{U}}=1.00$,\hspace{0.5cm}$\al=\pi/4$}}%
\put(4117,2611){\makebox(0,0)[r]{$b^{\lrp{U}}-a^{\lrp{U}}=0.75$,\hspace{0.5cm}$\al=\pi/4$}}%
\put(4117,2711){\makebox(0,0)[r]{$b^{\lrp{U}}-a^{\lrp{U}}=1.00$,\hspace{0.5cm}$\al=\pi/15$}}%
\put(4117,2811){\makebox(0,0)[r]{$b^{\lrp{U}}-a^{\lrp{U}}=0.75$,\hspace{0.5cm}$\al=\pi/15$}}%
\put(2565,50){\makebox(0,0){Masa del Higgs $h^0$ (GeV)}}%
\put(100,1612){%
\special{ps: gsave currentpoint currentpoint translate
270 rotate neg exch neg exch translate}%
\makebox(0,0)[b]{\shortstack{$Br\lrp{t\to ch^0}$}}%
\special{ps: currentpoint grestore moveto}%
}%
\put(4530,200){\makebox(0,0){ 160}}%
\put(4093,200){\makebox(0,0){ 155}}%
\put(3657,200){\makebox(0,0){ 150}}%
\put(3220,200){\makebox(0,0){ 145}}%
\put(2783,200){\makebox(0,0){ 140}}%
\put(2347,200){\makebox(0,0){ 135}}%
\put(1910,200){\makebox(0,0){ 130}}%
\put(1473,200){\makebox(0,0){ 125}}%
\put(1037,200){\makebox(0,0){ 120}}%
\put(600,200){\makebox(0,0){ 115}}%
\put(550,2924){\makebox(0,0)[r]{ 0.0024}}%
\put(550,2487){\makebox(0,0)[r]{ 0.002}}%
\put(550,2049){\makebox(0,0)[r]{ 0.0016}}%
\put(550,1612){\makebox(0,0)[r]{ 0.0012}}%
\put(550,1175){\makebox(0,0)[r]{ 0.0008}}%
\put(550,737){\makebox(0,0)[r]{ 0.0004}}%
\put(550,300){\makebox(0,0)[r]{ 0}}%
\end{picture}%
\endgroup
 
\end{center}
\caption{Branching ratio $B\left(t\to ch^0\right)$ for the process $t\to
ch^0 $ in the Fritzsch ansatz and using different values of the Yukawa
coupling parameters and ${\protect\alpha} $.}
\label{Fig1}
\end{figure}

%\subsection{Decays of the light neutral Higgs boson $h^0$}
Now, turn on the attention to the lightest neutral Higgs boson $h^0$ decays.
The $h^0$ decays to the quark pairs $d\overline{{\ b}}$, $s\overline{{\ b}}$, $d\overline{{\ s}} $ and $b\overline{{\ b}} $ are considered. In the
framework of the SM, the channel $h^0_{SM}\to b \overline{{\ b}} $ is the
most studied channel because it plays a central role for its possible
detection in the range $120-140$ GeV and its signature will be clear enough.
On the other hand, decays like $h^0 \to s\overline{{\ b}} $ are interesting
to study because FCNC in the Higgs sector would be a clear evidence of
physics beyond the SM. In order to look for differences, the ratio $r_{bb}=\frac{%
\Gamma\left(h^0\to b\bar{b}\right)_A}{\Gamma\left(h^0\to b\bar{b}\right)_{SM}%
}$ between the decay widths $\Gamma\left(h^0\to b\bar{b}\right)_A$ and $\Gamma\left(h^0\to b\bar{b}\right)_{SM}$ is defined. Since for all ansatze
considered in this work, the Yukawa couplings of the neutral Higgs boson $%
h^0 $ and the quark pair $b\overline{{\ b}} $ have the same intensity, the
ratio $r_{bb}$ has the same value for all ansatze. However, the ratio $r_{bb}$
gets the maximum value when $\cot {\alpha} =-a^D$, then for ${\alpha} =-\pi/4$
and $a^D=1$, the ratio is $2$, its mean that the decay width for these ansatze
can be twice the decay width predicted by the SM.\\

Another possible definition is the ratio, ${\gamma} ^{\left(A\right)}_{sb}=%
\frac{{\Gamma} \left(h^0\to s\overline{{\ b}} \right)_A}{{\Gamma}
\left(h^0\to b\overline{{\ b}} \right)_A}$ . From Table \ref{tab:table2d},
the relative decay widths for the process $h^0\to s\overline{{\ b}} $
predicted by the Fritzsch ansatz are at least one order of magnitude bigger
than the corresponding to the assignments type A and B of the Du-Xing
ansatz. This is due to the fact that the intensity of the Yukawa couplings
to the third generation fermions for the Fritzsch ansatz is stronger than
the one corresponding to the assignments type A and B of the Du-Xing ansatz,
which implies that the Fritzsch ansatz leads to a bigger probability of
finding the decay processes $h^0\to s\overline{{\ b}} $ than the obtained by
these assignments of the Du-Xing ansatz.\newline

\begin{table}[h]
\begin{tabular}{|c|c|c|c|}
\hline
& ${\alpha }=\frac{\pi }{3}$ & ${\alpha }=\frac{\pi }{18}$ & ${\alpha }=%
\frac{4\pi }{9}$ \\ \hline\hline
${\gamma }_{sb}^{\left( FA\right) }$ & $7.29\times 10^{-2}$ & $2.88\times
10^{-2}$ & $8.95\times 10^{-4}$ \\ \hline
${\gamma }_{sb}^{\left( DXA\right) }$ & $7.10\times 10^{-4}$ & $5.61\times
10^{-4}$ & $4.71\times 10^{-5}$ \\ \hline
${\gamma }_{sb}^{\left( FDXA\right) }$ & $7.10\times 10^{-4}$ & $5.61\times
10^{-4}$ & $4.71\times 10^{-5}$ \\ \hline
${\gamma }_{sb}^{\left( DXFA\right) }$ & $7.29\times 10^{-2}$ & $2.88\times
10^{-2}$ & $8.95\times 10^{-4}$ \\ \hline
${\gamma }_{sb}^{\left( \tilde{X}A\right) }$ & $1.92\times 10^{-3}$ & $%
1.51\times 10^{-3}$ & $4.71\times 10^{-5}$ \\ \hline\hline
\end{tabular}%
\caption{Relative decay width ${\protect\gamma }_{sb}^{\left( A\right) }=%
\frac{\Gamma \left( h^{0}\rightarrow s\bar{b}\right) _{A}}{\Gamma \left(
h^{0}\rightarrow b\bar{b}\right) _{A}}$ for the process $h^{0}\rightarrow s%
\overline{{\ b}}$ in the different ansatze with $M_{h^{0}}=120GeV$ and the
Yukawa coupling coefficients of the order of one.}
\label{tab:table2d}
\end{table}

%subsection{Charged Higgs decays}
Now, about the charged Higgs boson decays, two cases are considered: a
charged Higgs boson lighter than the top quark, $m_H=150$GeV and a heavier
charged Higgs boson $m_H=250$GeV. In the first case, the top quark could
decay into a charged Higgs boson and it will be an alternative to the usual
channel $t \to W^+ b$. Taking a light charged Higgs boson, the interesting
channels to detect a Higgs boson would be $H^+ \to c \bar{s}$, $H^+ \to c%
\bar{b}$ and $t \to H^+ b$. The decay widths were evaluated using phases
different of zero, $\psi_D=\theta_D=\frac{\pi}{18}$ but we also did it with
phases equal to zero and the results do not change in order of magnitude. In
table \ref{tab:table4} the results are shown, notice that the channel $H^+
\to c \bar{b}$ could compete with the searched channel $H^+ \to c \bar{s}$,
however in the FNA and NTA ansatze the $c \bar{s}$ channel is bigger than
the $c \bar{b}$ channel. About the option $t \to H^+ b$ the branching
fraction is of the order of $10^{-1}$ in the different ansatze considered.

In the case of a charged Higgs boson heavier than the top quark, $%
\Gamma(H^+\to t \bar{b})$, ${\Gamma} (H^+ \to t \bar{s})$, ${\Gamma} (H^+ \to
t \bar{d})$, $\Gamma(H^+ \to c\bar{s})$ and $\Gamma(H^+ \to c\bar{b})$ are
evaluated and they are shown in Table \ref{tab:table5}. It is worth to
notice that the most important channel would be $H^+ \to t \bar{b}$ except for the
FNA where it is $H^+ \to c \bar{s}$ .

\begin{table}[h]
\begin{tabular}{|c|c|c|c|c|c|c|c|}
\hline
${\Gamma }(GeV)$ & FA & DXA & FDXA & DXFA & $\tilde{X}$A & FNA & NTA \\
\hline
${\Gamma }\left( H^{+}\rightarrow c\bar{s}\right) $ & $4.83\times 10^{-4}$ &
$3.77\times 10^{-5}$ & $6.11\times 10^{-5}$ & $3.34\times 10^{-5}$ & $%
3.49\times 10^{-5}$ & $2.81$ & $3.72\times 10^{-5}$ \\ \hline
${\Gamma }\left( H^{+}\rightarrow c\bar{b}\right) $ & $1.04\times 10^{-2}$ &
$3.87\times 10^{-5}$ & $1.06\times 10^{-2}$ & $1.12\times 10^{-4}$ & $%
4.41\times 10^{-5}$ & $5.28\times 10^{-3}$ & $3.15\times 10^{-6}$ \\ \hline
$B\left( t\rightarrow H^{+}b\right) $ & $1.22\times 10^{-1}$ & $1.24\times
10^{-1}$ & $1.24\times 10^{-1}$ & $1.24\times 10^{-1}$ & $1.24\times 10^{-1}$
& $1.24\times 10^{-1}$ & $1.24\times 10^{-1}$ \\ \hline\hline
\end{tabular}%
\caption{The decay widths ${\Gamma }\left( H^{+}\rightarrow c\bar{s}\right) $%
, ${\Gamma }\left( H^{+}\rightarrow c\bar{b}\right) $ and the fraction $%
B(t\rightarrow H^{+}b)$ using the mass matrices ansatze discussed and
setting up $M_{H^{+}}=150GeV$, $\protect\psi _{D}=\protect\theta _{D}=\frac{%
\protect\pi }{18}$}
\label{tab:table4}
\end{table}

\begin{table}[h]
\begin{tabular}{|c|c|c|c|c|c|c|c|}
\hline
${\Gamma }(GeV)$ & FA & DXA & FDXA & DXFA & $\tilde{X}$A & FNA & NTA \\
\hline
${\Gamma }\left( H^{+}\rightarrow c\bar{s}\right) $ & $8.06\times 10^{-4}$ &
$6.29\times 10^{-5}$ & $1.02\times 10^{-4}$ & $5.57\times 10^{-5}$ & $%
5.82\times 10^{-5}$ & $4.68$ & $6.21\times 10^{-5}$ \\ \hline
${\Gamma }\left( H^{+}\rightarrow c\bar{b}\right) $ & $1.74\times 10^{-2}$ &
$6.45\times 10^{-5}$ & $1.76\times 10^{-2}$ & $1.87\times 10^{-4}$ & $%
7.35\times 10^{-5}$ & $8.80\times 10^{-3}$ & $5.26\times 10^{-6}$ \\ \hline
${\Gamma }\left( H^{+}\rightarrow t\bar{b}\right) $ & $1.32$ & $1.34$ & $%
1.34 $ & $1.34$ & $1.34$ & $1.34$ & $1.34$ \\ \hline
${\Gamma }\left( H^{+}\rightarrow t\bar{s}\right) $ & $6.12\times 10^{-2}$ &
$4.57\times 10^{-4}$ & $7.71\times 10^{-3}$ & $2.64\times 10^{-2}$ & $%
1.38\times 10^{-3}$ & $2.23\times 10^{-3}$ & $1.38\times 10^{-3}$ \\ \hline
${\Gamma }\left( H^{+}\rightarrow t\bar{d}\right) $ & $2.69\times 10^{-3}$ &
$2.71\times 10^{-5}$ & $6.44\times 10^{-4}$ & $1.41\times 10^{-3}$ & $%
6.90\times 10^{-5}$ & $2.03\times 10^{-4}$ & $7.24\times 10^{-5}$ \\
\hline\hline
\end{tabular}%
\caption{The decay widths ${\Gamma }\left( H^{+}\rightarrow c\bar{s}\right) $%
, ${\Gamma }\left( H^{+}\rightarrow c\bar{b}\right) $, ${\Gamma }%
(H^{+}\rightarrow t\bar{b})$, ${\Gamma }(H^{+}\rightarrow t\bar{s})$ and ${%
\Gamma }(H^{+}\rightarrow t\bar{d})$ in the different ansatze using $%
M_{H^{+}}=250GeV$ and $\protect\psi _{D}=\protect\theta _{D}=\frac{\protect%
\pi }{18}$.}
\label{tab:table5}
\end{table}

%\newpage

\section{\label{sec:level4}Conclusions}

The quark mass matrices ansatze proposed by Fritzsch, Du-Xing and
Fukuyama-Nishiura \cite{Fritzsch, Xing, Fukuyama-Nishiura} have been
reviewed in the framework of the general two Higgs doublet model and the
corresponding Yukawa matrices in the flavor basis. For these ansatze, the
numerical values of the CKM matrix elements and their experimental values
have been compared, obtaining that the Fukuyama-Nishiura ansatz leads to the
best agreement with the experimental results. The CKM matrix elements by
combining the Fritzsch and the Du-Xing ansatze for the up-type and down-type
quarks in different ways have also been obtained. For the CKM matrix
obtained by using the Fritzsch and the Du-Xing ansatze for the $U$-type and $%
D$-type quarks respectively a better agreement with the experimental CKM
matrix elements is gotten than the resulting when other Fritzsch and Du-Xing
ansatze combinations are used. Moreover, the CKM matrix by using two
different assignments of the Du-Xing ansatz for the $U$-type and $D$-type
quarks has been computed, obtaining a very good consistency between the
magnitudes of six of their elements and their experimental values. In the
concerning to the Fukuyama-Nishiura ansatz, the Yukawa coupling matrices in
the mass basis for both types of quarks have been computed by using two
different assignments for the $U$-type and $D$-type quarks. In this case,
vanishing entrances of the $\xi^U$ and $\xi^D$ matrices are obtained, they
are implying the absence of the flavor changing neutral currents involving
terms of the form $h^0\overline{{\ u}} c$, $h^0\overline{{\ c}} t$, $h^0%
\overline{{\ d}} b$, $h^0\overline{{\ s}} b$. Finally, a new type of ansatz
has been proposed where the FCNCs involving the top quark have vanished
completely and an excelent agreement with the CKM experimental elements is
gotten, with the exception of the $\bigl|K_{ub}\bigr|$, $\bigl|K_{cb}\bigr|$ and $\bigl|K_{td}\bigr|$ elements. Results are shown in Tables \ref{tab:table2} and \ref{tab:table3}.

On the other hand, a discussion about the phenomenology of the two body
decays of the lightest Higgs boson, the top quark and the charged Higgs
boson using the different mass matrices ansatze in the framework of the 2HDM
type III is presented in section \ref{sec:level3} and results are shown in
tables \ref{tab:table2d}, \ref{tab:table4} and \ref{tab:table5}. For the
lightest neutral Higgs boson, the decays $h^0 \to b\bar{b}$ and $h^0 \to b
\bar{s}$ are interesting. In the different ansatze considered the channel $b
\bar{b}$ can be enhanced respect to the SM using appropriate values of the
Yukawa parameters. And the channel $b \bar{s}$ would be more important in
the Fritzsch ansatz than in the other ansatze. About the charged Higgs
decays two options have been explored: a lighter charged Higgs boson an a
heavier one than top quark. Taking a lighter charged Higgs boson is possible
to get a bigger decay width for $H^+ \to c \bar{b}$ than the decay width for
$H^+ \to c \bar{s}$ depending on the used ansatz. And in the case of a
heavier charged Higgs boson than the top quark, if the Fukuyama-Nishiura
ansatz is used, the most important channel would be $H^{+}\rightarrow c%
\overline{{s}}$. Also in the case of a heavier charged Higgs boson than the
top quark, the channels with a top quark in the final state would be
relevant in a eventual search of the charged Higgs boson.

We acknowledge to R. Diaz for useful discussions. This work has been
supported by COLCIENCIAS and Fundaci\'on Banco de la Republica.

\appendix
\section{}
\subsection{CKM mixing matrix}

The corresponding standard parametrization of the CKM mixing matrix is \cite%
{PDG},
\begin{equation}
K=%
\begin{pmatrix}
c_{12}c_{13} & s_{12}c_{13} & s_{13}e^{-i\delta} \\
-s_{12}c_{23}-c_{12}s_{23}s_{13}e^{i\delta} &
c_{12}c_{23}-s_{12}s_{23}s_{13}e^{i\delta} & s_{23}c_{13} \\
s_{12}s_{23}-c_{12}c_{23}s_{13}e^{i\delta} &
-c_{12}s_{23}-s_{12}c_{23}s_{13}e^{i\delta} & c_{23}c_{13}%
\end{pmatrix}
\label{b350}
\end{equation}
with $c_{ij}=\cos\theta_{ij}$, $s_{ij}=\sin\theta_{ij}$ con $i,j=1,2,3$, and
$\delta$ the CP violating phase.

\subsection{Quark masses}

In the numerical computations, we used the central values of
the following quark masses at the energy scale of the order of $m_{Z}$ \cite{PDG}:
\begin{equation}
m_{u}\left( m_{Z}\right) =1.64\pm 0.40MeV,\hspace{1cm}m_{c}\left(
m_{Z}\right) =620\pm 30MeV,\hspace{1cm}m_{t}\left( m_{Z}\right) =172.7\pm
2.9GeV \label{b22}
\end{equation}%
\begin{equation}
m_{d}\left( m_{Z}\right) =2.92\pm 0.60MeV,\hspace{1cm}m_{s}\left(
m_{Z}\right) =55.56\pm 8.00MeV,\hspace{1cm}m_{b}\left( m_{Z}\right) =2.85\pm
0.18GeV \label{b23}
\end{equation}

%\newpage

\renewcommand{\refname}{References}

\section*{References}
\vspace{-1.3cm}
\bibliographystyle{plain}
\bibliography{apssamp}

\end{document}